\RequirePackage{fix-cm}
\documentclass[smallextended]{svjour3}       

\smartqed  
\usepackage[T1]{fontenc}
\usepackage[utf8]{inputenc}
\usepackage{amsfonts}
\usepackage{booktabs}
\usepackage{graphicx}
\usepackage{colortbl}
\usepackage{pbox}
\usepackage[numbers]{natbib}
\usepackage[most]{tcolorbox}
\usepackage{listings}
\usepackage{hyperref}
\usepackage{url}
\usepackage{multirow}
\usepackage{fourier}
\tcbset{
  customsummarybox/.style={
    colback=gray!10,       
    colframe=gray!50,      
    boxrule=0.5mm,         
    arc=5mm,               
    width=\linewidth,      
    left=1mm,              
    right=1mm,             
    top=1mm,               
    bottom=1mm,            
    sharp corners=downhill,
  }
}

\begin{document}

\title{Empirical Studies of Parameter Efficient Methods for Large Language Models of Code and Knowledge Transfer to R\thanks{This research is supported by a grant from Natural Sciences and
Engineering Research Council of Canada RGPIN-2019-05175 and Mitacs Accelerate award, 2024.}
}

\titlerunning{Parameter-Efficient Fine-Tuning in Software and Knowledge Transfer to R}        

\author{Amirreza Esmaeili \and 
        Iman Saberi       \and
        Fatemeh Fard    
}


\date{Received: date / Accepted: date}

\institute{Amirreza Esmaeili \at
              University of British Columbia \\
              3333 University Way, Kelowna, BC V1V 1V7, Canada \\
              \email{a.esmaeili@ubc.ca}           
           \and
           Iman Saberi \at
              University of British Columbia \\
              3333 University Way, Kelowna, BC V1V 1V7, Canada \\
              \email{iman.saberi@ubc.ca}           
           \and
           Fatemeh Fard \at
              University of British Columbia \\
              3333 University Way, Kelowna, BC V1V 1V7, Canada \\
              \email{fatemeh.fard@ubc.ca}           
}

\maketitle

\begin{abstract} 
Parameter Efficient Fine-Tuning (PEFT) methods are proposed as an alternative fine-tuning approach for Large Language Models (LLM) to minimize high training costs. 
While prior research demonstrates the effectiveness of PEFT methods in knowledge transfer using smaller language models, their application to larger LLMs, particularly in low-resource and unseen programming languages such as R, remains under-explored.
In this work, we evaluate PEFT methods, LoRA, Compacter, and IA$^3$ on LLMs 
for code summarization and generation, with a particular emphasis on knowledge transfer to R as an unseen under-explored target language. 
Our experiments reveal that LoRA consistently outperforms Compacter and IA$^3$ in all settings, while Compacter offers significant resource efficiency with minimal performance trade-offs. Additionally, we find that the number of trainable parameters has a greater influence on the functional accuracy of the generated code than PEFT architecture.
Our study can direct future research in developing code intelligent tasks for unseen languages including R, as well as the choice of PEFT methods for knowledge transfer, especially when balancing the computational cost and performance.

  \keywords{Parameter Efficient Fine Tuning, Code Large Language Models, Knowledge Transfer, Unseen Languages, Low-Resource Language, R.}
\end{abstract}

\section{Introduction}\label{sec:introduction}
Large Language Models (LLMs) such as T5 model \cite{t5-paper}, Llama2 \cite{llama2-touvron2023llama}, GPT-3 \cite{gpt3-NEURIPS2020_1457c0d6}, CodeT5 \cite{codet5-wang-etal-2021-codet5} and CodeLlama \cite{codellama-roziere2023code} have achieved the state-of-the-arts performance in many natural language and software engineering tasks, including code summarization~\cite{codet5-wang-etal-2021-codet5} and code generation \cite{codet5-wang-etal-2021-codet5, codellama-roziere2023code}. 
Considering the heavy computational cost of the models, Parameter Efficient Fine-Tuning (PEFT) methods are used to tune a small percentage of the models' parameters to a downstream task. 
There are various studies ~\cite{saberi2023multilingual,saberi2023utilization, weyssow2023exploring, ase2023, tse2023, icse2023} that explore the application of PEFT methods such as Adapter \cite{adapter-houlsby2019parameter}, LoRA \cite{lora-hu2022lora}, Prompt Tuning \cite{prompt-lester-etal-2021-power} and IA$^3$ \cite{ia3-liu2022few} in software engineering. 
Some of these works \cite{rel1-Compre, ase2023, goel2022cross} employ smaller models (less than $1$B parameters) such as CodeBERT \cite{codebert-feng-etal-2020-codebert} and GraphCodeBERT \cite{graphcodebert-guo2020graphcodebert}, while others are specifically targeting larger models (more than $1$B parameters)~\cite{weyssow2023exploring} in a diverse set of software engineering tasks such as code search, code summarization, clone detection, defect detection, and code generation.

Previous research has shown the advantage of using PEFT methods in knowledge transfer from natural language to code~\cite{goel2022cross, saberi2023utilization} using smaller models CodeBERT \cite{codebert-feng-etal-2020-codebert} and GraphCodeBERT \cite{graphcodebert-guo2020graphcodebert}, and to low-resource languages, even outperforming full fine-tuning of the language models~\cite{saberi2023utilization}.
Other studies \cite{rel8-transfer, cassano2023knowledge, saberi2023multilingual} cover knowledge transfer to the domain of \emph{low-resource} and unseen languages, where training data is limited and the base model has not encountered the target language.
However, there is little known about how PEFT methods perform when they are applied to LLMs, especially for knowledge transfer from natural language to code and also how they perform on R. 
Moreover, within all studies on PEFT methods in the software engineering domain to this date, the lack of an in-depth investigation of Compacter \cite{compacter-mahabadi2021compacter} is notable. Proposed by \citet{compacter-mahabadi2021compacter}, Compacter further improves the parameter efficiency of Adapters \cite{adapter-houlsby2019parameter} by incorporating low-rank matrices and Kronecker multiplications.

Thus, we aim to assess the performance of the PEFT methods to determine their effectiveness in knowledge transfer from natural language to code, and whether they can effectively serve as substitutes for conventional full fine-tuning, specifically in unseen and low-resource scenarios.
For the unseen and low-resource programming language, we choose R.  
First, given that R is rarely included in the training of common LLMs, it serves as an ideal target language to assess the performance of PEFT methods on low-resource and unseen programming languages.
Second, R is a widely adopted programming language with a large community and ecosystem, especially known for its applications in statistical computing and data visualization. While not as popular as languages such as Python or Javascript, R is ranked 18th on the TIOBE index\footnote{https://www.tiobe.com/tiobe-index} in November 2024 and 21st on Stack Overflow's 2024 Developer Survey\footnote{https://survey.stackoverflow.co/2024/technology}. Choosing R as a target language can ultimately lead to further adoption and a better understanding of the models' capabilities in this language. 

We investigate T5 \cite{t5-paper} and Llama2 \cite{llama2-touvron2023llama}, commonly employed open-source natural language LLMs (that we refer to as general-LLM), and code large language models (that we refer to as code-LLMs), CodeT5 \cite{codet5-wang-etal-2021-codet5} and CodeLlama \cite{codellama-roziere2023code}, fine-tuned with three PEFT techniques: Compacter, LoRA, and IA$^3$. 
Due to the large number of experiments and computational restrictions, we consider code summarization and code generation, which are two well-known tasks in software and have been studied extensively in recent years \cite{codet5-wang-etal-2021-codet5,codellama-roziere2023code,graphcodebert-guo2020graphcodebert,codebert-feng-etal-2020-codebert}.

Through extensive empirical experiments, we demonstrate that LoRA consistently outperforms both Compacter and IA$^3$ for code summarization and code generation. This includes adapting from natural language to code on general-LLMs, from code to natural language on code-LLMs, and transferring knowledge to R as an unseen language. Furthermore, we show that while Compacter incurs a relatively small performance trade-off, it offers significant resource efficiency, making it an attractive choice in scenarios where minimizing computational requirements is a priority. Finally, our findings highlight that the number of trainable parameters has a greater impact on functional accuracy in code generation than the specific architecture of PEFT methods.

Our main contributions are an in-depth analysis of PEFT comparison using LLMs, domain adaptation from natural language and code-specific LLMs to code-related tasks, and adapting LLMs to unseen language, R. 
We make all the scripts available to support open science. The link is found in Section \ref{sec:data-ack}.

Employing PEFT methods as an alternative fine-tuning process in code-related tasks can notably decrease both training time and resource usage, ultimately reducing the cost and environmental side-effects of training large language models, making the LLMs more accessible in cases where computational resources are scarce. 
The focus on the low-resource language of R helps in research using LLMs to automate software engineering tasks for a wider community of developers. 
Our research provides interesting results in this direction and can guide future research for code intelligence for R, as well as using the studied PEFT methods, specifically LoRA and Compacter with LLMs.  

\textbf{Note.} \textit{Please note that this work is the submission of our accepted Registered Report in MSR 2024 \href{https://2024.msrconf.org/track/msr-2024-registered-reports?\#event-overview}{Registered Report Track}, titled ''Empirical Studies of Parameter Efficient Methods for Large Language Models of Code and Knowledge Transfer to R". The link to the registered report can be found \href{https://arxiv.org/abs/2405.01553}{here}.}

\textbf{Deviations from Registered Report.}
There are small deviations from the registered report in this manuscript as follows. We discuss the details of the methodology and each process in Section~\ref{sec:study-design}.
\begin{itemize}
    \item We adopted the SPP dataset for Python code generation instead of the CoNaLa dataset, which offers better compatibility with the dataset used for R code generation and benchmarks used to assess code generation quality, in addition to a larger filtered training data sample size, $28,182$ compared to $2,379$ samples, which is more beneficial for large LLMs.
    \item We opted out of implementing a new R code generation benchmark, as a high-quality alternative came into existence that adequately aligns with the HumanEval Python benchmark, resulting in more consistent assessments.
    \item We initially proposed two scenarios in RQ3, the first one being excluding the target unseen language from the model during training, i.e., excluded during the pre-training and fine-tuning phase. We eliminated this scenario due to two reasons. We conducted initial experiments with some of the models and languages and we observed small differences between these models and the base models with no fine-tuning. More importantly, due to several configurations for the models and PEFT methods that we considered to ensure the reliability of the results, the number of experiments would increase significantly. Therefore, due to computational resources and our initial experiments, we eliminated this scenario.
    \item We utilized a zero-shot setting as the baseline for Llama models in place of full fine-tuning, as the substantial time and computational resources required for fine-tuning of such models were beyond our computational capacity.
    \item We included IA$^3$ as the third PEFT method with limited experiments, which shed light on this novel fine-tuning technique and better elaborate the attributes of other PEFT methods.
\end{itemize}

\section{Background}\label{sec:background}
\subsection{Pre-trained Language Models}

Traditionally, researchers were required to collect large amounts of task-related annotated data to train a model. 
With the advent of the Transformer architecture \cite{transformer} and pre-training \cite{pretraining, pretraining-peters-etal-2018-deep, pretraining-radford2018improving}, language models are introduced; models that are pre-trained on a vast amount of unlabelled data enabling the model to obtain commonsense information about a language or multiple languages. Then, these models are fine-tuned on supervised datasets for different downstream tasks. NLP researchers have proposed several pre-trained language models such as BERT \cite{bert-Devlin2019BERTPO}, T5 \cite{t5-paper}, and GPT-3 \cite{gpt3-NEURIPS2020_1457c0d6}, serving as a checkpoint of language knowledge ready for further training. 

Early contributions have shown promising results by utilizing models pre-trained on code and natural language in the SE domain, and achieved state-of-the-art results for several tasks such as vulnerability detection \cite{nlp-code-fu-VulRepair}, code classification \cite{nlp-code-Yang2021DeepSCCSC}, code search \cite{li2022coderetriever}, code summarization \cite{codet5-wang-etal-2021-codet5} and code generation \cite{codet5+-wang2023codet5+}.
Since the introduction of CodeBERT~\cite{codebert-feng-etal-2020-codebert}, several lanauge models are introduced in the SE domain, GraphCodeBERT~\cite{graphcodebert-guo2020graphcodebert}, CodeT5 \cite{codet5-wang-etal-2021-codet5}, CoTexT~\cite{cotext-phan-etal-2021-cotext}, AlphaCode~\cite{alphacode}, CodeX~\cite{humaneval-codex-Chen2021EvaluatingLL} and CodeLlama \cite{codellama-roziere2023code}. 

\subsection{Parameter Efficient Fine-Tuning}

Recent studies have demonstrated the feasibility of modifying or adding a relatively small subset of parameters \cite{adapter-houlsby2019parameter}, in contrast with the standard full fine-tuning approach, which involves updating all parameters of a language model.
Parameter-efficient fine-tuning significantly reduces memory and storage requirements during model training. State-of-the-art PEFT methods have demonstrated the capability to reach the performance achieved by fully fine-tuning a model by modifying a small percentage of the model's parameters.
Additionally, PEFT methods can achieve better accuracy than fully fine-tuning a model in low-resource settings \cite{saberi2023multilingual, low-res-related}, where labelled data for fine-tuning is scarce.

PEFT encompasses several methods, such as inserting small trainable feed-forward layers into a fixed pre-trained Transformer-based model~\cite{adapter-houlsby2019parameter}, only updating the bias parameters while freezing the rest of a model's parameters~\cite{bitfit-ben-zaken-etal-2022-bitfit}, decomposing weight matrices into approximate low-rank matrices~\cite{lora-hu2022lora}, introducing Kronecker multiplication with weight matrices and utilizing low-rank matrices~\cite{compacter-mahabadi2021compacter}, concatenating embedding weight matrices to fixed models~\cite{prefix-li-liang-2021-prefix, prompt-lester-etal-2021-power}, and adding efficient element-wise multiplication to the feed-forward and attention layers of models to rescale weight matrices~\cite{ia3-liu2022few}.
In this work, we will study LoRA \cite{lora-hu2022lora}, Compacter \cite{compacter-mahabadi2021compacter} and IA$^3$ \cite{ia3-liu2022few}. Though LoRA was previously experimented with within SE, the value of Compacter and IA$^3$ is not well-known for SE tasks. 

\textbf{LoRA.}
Proposed by~\citet{lora-hu2022lora}, LoRA draws inspiration from IntrinsicSAID \cite{intrinsicsaid-aghajanyan-etal-2021-intrinsic} and introduces a more straightforward method for executing low-rank fine-tuning. In LoRA, the parameter update process for a weight matrix $\delta W$ involves decomposing it into a product of two low-rank matrices $W_A$ and $W_A$:

\begin{equation}
\begin{aligned}
\delta W = W_A W_B
\end{aligned}
\end{equation}
Where $W_A \in \mathbb{R}^{in \times r}$ and $W_B \in \mathbb{R}^{r \times out}$.

Notably, LoRA maintains the pre-trained model parameters in a frozen state, allowing only the $W_A$ and $W_B$ matrices to be trainable. The scaling factor is typically constant and often set to $1/r$.
Post-training, these decomposed matrices—$W_A$ and $W_B$—can be seamlessly integrated back into the original matrix $W$ by a straightforward addition of the matrix product $W_AW_B$. 

This method showcases notable efficacy in achieving improved performance while employing a simplified approach to low-rank fine-tuning, showcasing its applicability across extensive model architectures.

\textbf{Compacter.}
\citet{compacter-mahabadi2021compacter} proposed Compacter, which adopts a unique approach, leveraging the Kronecker product, low-rank matrices, and parameter sharing across layers to create adapter weights. Within each adapter, every $W$ parameter is constructed as a sum of $n$ Kronecker products of a set of smaller dimension matrices $A_i$ and $B_i$:

\begin{equation}
\begin{aligned}
\hat{W} = \sum_{i=0}^{n} A_i \otimes B_i
\end{aligned}
\end{equation}

Where $\hat{W} \in \mathbb{R}^{k \times d}$, $\quad A_i \in \mathbb{R}^{n \times n}$, $\quad B_i \in \mathbb{R}^{\frac{k}{d} \times \frac{n}{d}}$.

This results in forming a linear layer referred to as the parametrized hypercomplex multiplication (PHM) layer \cite{phm-Zhang2021BeyondFL}.
Expanding on this, Compacter refines its methodology by parametrizing $B_i$ similar to LoRA, with the expression $B_i = B_i^{down} B_i^{up}$. Here, all matrices maintain a maximum rank of $r$. Additionally, $A_i$ matrices are shared across all adapter layers to enhance parameter efficiency. 

\textbf{IA$^3$.}
In a recipe dubbed as \emph{T-Few}, \citet{ia3-liu2022few} proposed IA$^3$, which introduces three trainable vectors to the keys and values of the attention layers and the feed-forward layers of a transformer-based language model. These vectors rescale the parameters of their respective parameter modules during the fine-tuning phase by operating on an element-wise level.
Formally, the attention layers of an IA$^3$-infused model are transformed as:

\begin{equation}
\begin{aligned}
softmax(\frac{Q(l_k \odot K^T)}{\sqrt{d_k}})(l_v \odot V)
\end{aligned}
\end{equation}

and the feed-forward modules transformed as:

\begin{equation}
\begin{aligned}
W_2(l_{ff} \odot \gamma(W_1x)),
\end{aligned}
\end{equation}

where $\gamma$ is the non-linearity of the feed-forward modules. 
The simplicity of the element-wise multiplication operations employed by IA$^3$ introduces little computational overhead.
Despite its straightforward operations, IA$^3$ achieves significant performance, surpassing the human baseline on the RAFT benchmark \cite{raft-Alex2021RAFTAR}. This benchmark, which focuses on real-world and economically valuable problems, marked IA$^3$ as the first method to exceed the human baseline at the time of its publication. 

\section{Study Design}\label{sec:study-design}
In this section, we detail the experimental setup and study design, aiming to address the following research questions (RQ) in our investigation:\\

\textbf{RQ1:What is the effectiveness of applying PEFT methods for the selected tasks in code-LLMs?}

In this RQ, we compare the performance of applying LoRA, Compacter and IA$^3$ on our selected LLMs for multiple programming languages, compared to their respective baselines. 
The results of this experiment will shed light on the capabilities of each PEFT method, while also providing insight on the adaptability of each LLM with these methods.
As there are new architectures for each PEFT approach, and there are various architectures for LLMs, it is important to understand which methods could be applied to which models and their results. 

\textbf{RQ2: How do PEFT methods perform when used for transferring knowledge from natural language to code with LLMs? }

Previous studies~\cite{goel2022cross, saberi2023utilization} show that a specific PEFT method, called adapters, can help in transferring knowledge from natural language models to code-related tasks. They focused on the adaptability of smaller models, i.e., RoBERTa~\cite{roberta-liu2019roberta} in comparison to pre-trained programming language models, such as CodeBERT~\cite{codebert-feng-etal-2020-codebert}, and demonstrate that RoBERTa plus adapters can achieve on-par results compared to CodeBERT for code summarization. Given the resource-intensive nature of pretraining large language models on new domains (i.e., code), we aim to investigate whether the PEFT methods can achieve similar results when applied to LLMs that are not pre-trained on code. 
If successful, the results would hold promise for significantly mitigating the computational resources required for training code-LLM.

\textbf{RQ3: How do PEFT-tuned LLMs perform in transferring their learned knowledge to an unseen programming language?} 

Prior research \cite{unseen-Radford2019LanguageMA} suggests that language models may encounter performance limitations when confronted with languages that were not part of their training data. 
Consequently, we focus on evaluating LoRA's, Compacter's and IA$^3$'s capabilities for leveraging the knowledge of a code-LLM for adapting to a programming language that is not seen during their pre-training phase.
This experiment offers valuable insights into the extent of PEFT methods' abilities for leveraging the learned knowledge of programming languages and adapting them to new ones. 

\subsection{Approach}

In this study, we employ two base code-LLM, CodeT5 \cite{codet5-wang-etal-2021-codet5} and CodeLlama \cite{codellama-roziere2023code}, and their counterparts, i.e., T5 and Llama2, when LoRA, Compacter and IA$^3$ are added to them. 

To answer RQ1, we fine-tune the code-LLMs on the datasets of the two tasks, i.e., code summarization and code generation, separately. The details of the datasets are in Section~\ref{sec:data}. For each of the tasks, we only consider the mono-lingual data and report the results for each programming language separately. 
Models fine-tuned with PEFT methods will be compared to their respective baselines. To this end, we employ a fully fine-tuned CodeT5 as the baseline of the experiments on this model. However, for the CodeLlama, we use a pre-trained checkpoint of the model as the baseline since this model is nearly an order of magnitude larger than the T5 family of models, and fully fine-tuning is not feasible given our resource limitation. 

To answer RQ2, we use T5 \cite{t5-paper} and Llama2 \cite{llama2-touvron2023llama} models and add LoRA, Compacter and IA$^3$ to these models, as an alternative to fine-tuning them for code summarization and code generation tasks. Similar to RQ1, we conduct experiments for each programming language separately. Additionally, following RQ1, we use a fully fine-tuned T5 and a pre-trained checkpoint of Llama2 as baselines to ensure fairness and reduce computational costs.

For RQ3, we choose R as the target programming language for both code summarization and code generation tasks, as it has a wide community, but is an under-explored language for code intelligence. R is considered a low-resource programming language (i.e., it has limited training data available) and thus is chosen here~\cite{cassano2023knowledge}. Moreover, R is not used in the training dataset of CodeT5 \cite{codet5-wang-etal-2021-codet5} and CodeLlama \cite{codellama-roziere2023code}, serving as a suitable unseen programming language. This is confirmed as the training dataset of CodeT5 is publicly available. Additionally, R is not among the list of programming languages used in the training of CodeLlama, as stated by its authors.
To answer this RQ, we fine-tune T5, CodeT5, Llama2 and CodeLlama using the PEFT methods on code summarization and code generation for R. Following the previous RQs, we use fully fine-tuned T5 and CodeT5 and pre-trained checkpoints of Llama2 and CodeLlama as baselines. 
This experiment gives insight into how capable each model and PEFT method is when transferring the learned knowledge from other languages used in pre-training to R.

\subsection{Downstream Tasks} \label{sec:tasks}

To evaluate the models, we choose the two code summarization and code generation tasks. These tasks are studied in several works \cite{codet5-wang-etal-2021-codet5,codet5+-wang2023codet5+,codellama-roziere2023code,saberi2023utilization,alphacode} and are two of the tasks on CodeXGLUE benchmark\footnote{\url{https://microsoft.github.io/CodeXGLUE/}}. We choose these tasks as many of these are code-to-text and text-to-code tasks, showing the capabilities of the LLMs for both code and natural language comprehension. Additionally, in RQ2, we intend to investigate the knowledge transferability of the LLMs to code intelligence tasks. Therefore, choosing code-to-code tasks might not be the best option for T5 and Llama2 models, as there is no natural language involved in those tasks. Exploring the full spectrum of this knowledge transferability for various code intelligence tasks is beyond the scope of this study. The last reason for choosing these tasks is the scarcity of the available data for R, which prohibits exploring more tasks.

\textbf{Code Summarization.}
The primary objective of a code summarization task is to distill the functionality and purpose of a code snippet or function into a concise and understandable summary. 
Code summarization enables streamlining documentation generation processes and automating commit generation.
We train the models on the code summarization datasets. Then we test the model by feeding the code snippets as input sequences and evaluate the description comments as target sequences ($Code \longrightarrow Comment$).
We employ this task to gauge the models' capabilities to analyze the structure, logic, and functionality of the code snippets \cite{task-ahmad2020transformer}.

\textbf{Code Generation.}
Code generation task aims to transform natural language descriptions or instructions into executable code or programming language constructs. This task enables the automation of code writing based on human-provided specifications or requirements. By converting natural language descriptions into functional code snippets, code generation contributes to automating software development processes and aids in rapid prototyping or code scaffolding. 
To train a model for code generation, we provide natural language descriptions, along with additional desired context such as a specific function header, as input sequences and expect the corresponding code snippets as target sequences ($Comment \longrightarrow Code$).
This setup allows us to assess the models' proficiency in understanding and translating human-readable descriptions into executable code segments.

\subsection{Datasets and Benchmarks} \label{sec:data}

\textbf{CodeSearchNet Dataset.} 
CodeSearchNet \cite{CSN-husain2020codesearchnet}, curated by Microsoft Research, is a dataset of $2,326,976$ functions paired with corresponding explanation comments in six programming languages, namely, Go, Java, JavaScript, PHP, Python, and Ruby. 
CodeSearchNet is derived from a diverse set of real-world code repositories, including GitHub. We have opted for the utilization of CodeSearchNet {in RQ1 and RQ2} because it has been a subject of recent scholarly investigations in the realm of code summarization \cite{codebert-feng-etal-2020-codebert,graphcodebert-guo2020graphcodebert,codet5-wang-etal-2021-codet5,saberi2023multilingual}. Throughout this paper, we refer to the programming languages present in CodeSearchNet as the \emph{six seen languages}.

\textbf{SPP Dataset.} The SPP dataset \cite{spp_dataset} consists of approximately $450,000$ synthetic Python programming problems, each containing a task description in the form of a function header and a comment, one to three use-cases, a code solution, and one to three test cases, generated using the CodeGeeX-13B model. For our experiments, we utilized a subset of this data {in RQ1 and RQ2}: a verified split containing $30,000$ samples that have been de-duplicated and verified by a Python interpreter. This ensures the side effect of low-quality data is eliminated from the experiments given the unverified subset may contain syntactically incorrect code and multiple function definitions per sample, degrading the data quality.

\textbf{R Code Summarization Dataset.}
We use the $R_{combined}$ subset of the R dataset curated by ~\citet{Zhao2024DoCL}, which contains $32,671$, and $4,078$, and $4,084$ code and comment pairs for train, evaluation, and test, respectively. 

This dataset is collected by scraping active and public GitHub repositories containing R files, which have an open-source license. The dataset is further filtered to eliminate repositories that are personal or do not follow proper R packaging rules. The collection is parsed using the Tree-sitter \footnote{https://tree-sitter.github.io/tree-sitter} parser and R documentation tools to transpile it into a dataset containing R code snippets and natural language descriptions~\cite{Zhao2024DoCL}. This dataset has the same structure as CodeSearchNet and assures consistency among datasets when comparing the results of code summarization. We utilize this dataset for the R code summarization tasks in RQ3.

\textbf{MultiPL-T.}
MultiPL-T is an approach for curating semi-synthetic datasets for low-resource languages proposed by \citet{cassano2023knowledge}. This approach translates high-quality code snippets of high-resource languages such as Python into Julia, Lua, OCaml, R, and Racket using a variation of the StarCoderBase-15B model \cite{li2023starcoder}. The source code snippets consist of test-validated function bodies and descriptive document strings, making the target datasets ideal for code generation tasks. 
The R split of the MultiPL-T dataset contains $37,592$ pairs of natural language descriptions and R functions. We utilize this dataset for R code generation in RQ3 by splitting it into $35,338$, $376$, and $1,860$ samples for train, validation and testing, respectively.

\textbf{HumanEval.} Proposed by OpenAI, HumanEval \cite{humaneval-codex-Chen2021EvaluatingLL} is a Python benchmark consisting of 164 hand-written instructions and on average 7.7 unit tests for each instruction for code generation evaluation. HumanEval directly assesses the functionality correctness of generated code by testing each sample against a suite of unit tests hand-crafted for each instruction within the benchmark. This method of code generation evaluation mitigates reported performance inaccuracy caused by functionally correct but textually dissimilar output samples. We will use this benchmark to evaluate our trained model for code generation in Python.

\textbf{MultiPL-E.} 
As a translation system for code generation test-driven benchmarks, MultiPL-E \cite{multipl_e} is designed to translate sample benchmark functions and their associated unit tests into new arbitrary programming languages, similar to MultiPL-T. The authors have utilized MultiPL-E to translate the HumanEval \cite{humaneval-codex-Chen2021EvaluatingLL} and MBPP \cite{mbpp_Austin2021ProgramSW} benchmarks into $18$ new languages, including R. As the R counterpart of HumanEval for Python code generation, we employ the R translation of the HumanEval benchmark created using MultiPL-E for this study. This benchmark has a structure similar to the HumanEval benchmark, containing $161$ problems. This benchmark is used for R code generation evaluation in this study.

\subsection{Evaluation Metrics}

For evaluating code summarization, we follow previous research by adopting the smoothed BLEU-4~\cite{bleu-paper} as our evaluation metric~\cite{ahmad2021unified, codet5+-wang2023codet5+, sontakke2022code}.
Aligning with previous methodologies, we maintain consistency by averaging the BLEU scores to derive a comprehensive assessment of the generated summaries in comparison to the reference texts.

\textbf{BLEU.} The BLEU score \cite{bleu-paper} is a metric originally proposed for evaluating the quality of translations at the document level. 
This metric relies on precision and evaluates the geometric precision of n-grams between the generated text and the ground truth text. For a translation $T$, a reference $R$ and a range from $1$ to $N$ of n-grams, the BLEU score is calculated as follows:

\begin{equation} \label{formula:bleu}
\begin{aligned}
BLEU(N, T, R) = (\prod_{n=1}^{N} \frac{m_n}{l_n})^\frac{1}{N} \times BP(T, R)
\end{aligned}
\end{equation}

where $m_n$ is the number of matching n-grams between a translation and its reference, and $BP(T, R)$ associates a penalty to the score when the length of the translation is less than the reference using the following formula:

\begin{equation}
\begin{aligned}
BP(T, R) = min(1, exp(1 - \frac{len(T)}{len(R)}))
\end{aligned}
\end{equation}

As noted, the BLEU score is used for evaluating a prediction across the entire reference corpus. For our use case (i.e., evaluating a sentence level output against a ground truth sentence), a smoothed variant of the BLEU score is required, which is obtained by replacing $m_n$ in formula \ref{formula:bleu} with a small positive number $\epsilon$ if there are no matching n-grams (i.e. $m_n  = 0$).
For evaluating code summarization, we follow prior research~\cite{ahmad2021unified, codet5+-wang2023codet5+, sontakke2022code} by adopting the smoothed BLEU score with $N = 4$ (henceforth referenced as BLEU-4) as our evaluation metric.

\textbf{CodeBLEU.} The BLEU score, designed to evaluate text similarity between generated and reference text, possesses limitations when applied to assess code generation tasks due to its emphasis on linguistic resemblance rather than functional equivalence. Consequently, accurately functional yet linguistically dissimilar code may receive suboptimal BLEU scores, rendering it an inadequate metric for code evaluation.
For code generation, we employ CodeBLEU~\cite{ren2020codebleu} and EM@k.

CodeBLEU is proposed to validate the syntactic and semantic data flow accuracy inherent in code generation models. CodeBLEU stands as a composite measure that injects code syntax via abstract syntax trees (AST) and code semantics via data flow. Compared to the BLEU score, This metric serves as a more comprehensive assessment tool, facilitating the quantification and analysis of the multifaceted aspects associated with the correctness of generated code.
\begin{equation*}
    CodeBLEU = \alpha \cdot BLEU + \beta \cdot BLEU_{weight}
+ \gamma \cdot Match_{ast} + \delta \cdot Match_{df}
\end{equation*}
Where $BLEU$, $BLEU_{weight}$, $Match_{ast}$, and $Match_{df}$ are calculated based on standard BLEU, weighted n-gram match, syntactic AST match, and semantic dataflow match, respectively. $\alpha$, $\beta$, $\gamma$ and $\delta$ determine the the weighted combination of four parts.

\textbf{EM@k.} Exact Match pertains to a strict quantification textual prediction, where if a prediction for a sample exactly matches the ground truth (or one of several ground truths) associated with that sample, $EM = 1$, otherwise $EM = 0$. In the scope of this metric, even a single character discrepancy between the prediction and the ground truth will result in $EM = 0$. 
Formally, the average Exact Match score of $k\in\{1,N\}$ candidate samples generated by the model for a reference ground truth $r$ is computed as follows:

\begin{equation}
\begin{aligned}
EM@k = \frac{\Sigma^{k}_{i=1} EM(s_i, r)}{k}
\end{aligned}
\end{equation}

In this study, we employ EM@1 as part of our evaluations for code generation experiments.

\textbf{Pass@k.} Despite alleviating some of the BLEU score's issues when applied for code generation and its widespread use in the academic scene, CodeBLEU still possesses caveats \cite{humaneval-codex-Chen2021EvaluatingLL}. For instance, the CodeBLEU score for a dissimilar (both textual and AST) generated code compared to the ground truth will degrade, even if the dissimilar generated code does satisfy the requested functionality. To address these issues, OpenAI has proposed HumanEval \cite{humaneval-codex-Chen2021EvaluatingLL}, a Python benchmark dataset, as explained above. The evaluation metric associated with HumanEval is $Pass@k$, where if among the $n > k$ generated sample codes $c$ samples pass all unit tests corresponding to an instruction, the evaluation metric is obtained as follows:

\begin{equation}
\begin{aligned}
Pass@k := E [1 - \frac{\binom{n - c}{k}}{\binom{n}{k}}]
\end{aligned}
\end{equation}

{In all experiments, we use greedy decoding for generation, thus eliminating randomness from generations. Consequently, we chose $k=1$ with one generation per sample to calculate Pass@k and EM@k, as subsequent generations with the same input produce the same results.}

\textbf{Textual and Functionality Similarity in Code Generation. }
Code snippets generated by a model can be assessed based on textual similarity and functional correctness. A snippet with high textual similarity to the ground truth may still fail functionally, as even a single incorrect operator can compromise its functionality despite its resemblance to the original code.
The BLEU-4 score measures textual similarity between a sample and the ground truth, making it suitable for code summarization. However, it falls short in evaluating functionality for code generation. CodeBLEU addresses this limitation by incorporating the abstract syntax tree and data flow, offering a more in-depth functional assessment.

Taking these characteristics into account, the only metric that ensures the code is executable and correct is the Pass@K.
The HumanEval benchmark addresses these limitations by evaluating correctness through a predefined set of test cases focused solely on functionality, rather than textual similarity to the ground truth. Since the problems in this benchmark are uniquely crafted, a high Pass@k HumanEval score suggests that the model may perform well on out-of-distribution problems. For the code generation experiments in this study, we use all the mentioned metrics when available, considering both textual similarity and functionality correctness.

\subsection{Base Models} \label{sec:models}

This section covers the base LLMs used in this study. We refer to the code-specific models as \emph{code-LLM} and general-purpose natural language models as \emph{general-LLM}.

\textbf{CodeT5.} CodeT5~\cite{codet5-wang-etal-2021-codet5} is a code-LLM based on T5 \cite{t5-paper}, pre-trained on CodeSearchNet \cite{CSN-husain2020codesearchnet}, and additional C and C\# data. By considering the type of identifiers in code and utilizing an encoder-decoder architecture, CodeT5 achieves state-of-the-art performance in a wide range of code comprehension and generation tasks in the CodeXGLUE benchmark \cite{codexglue-lu2021codexglue}.

\textbf{CodeLlama.} CodeLlama~\cite{codellama-roziere2023code} is another state-of-the-art decoder-only code-LLM that comes in different variants and parameter budgets. CodeLlama is based on Llama2 \cite{llama2-touvron2023llama} with further training on an extensive corpus consisting of 500 billion tokens (1 trillion tokens for CodeLlama 70B) in the case of the foundation model. CodeLlama is trained on a proprietary dataset, encompassing programming languages such as Java, Javascript, PHP, Python, C++, and C\#, among others.

 \textbf{T5.} T5~\cite{t5-paper} was proposed as a Text-to-Text Transformer, a unified model used for Transfer learning. The authors introduced a unified framework that converted all text-based problems into a text-to-text format. 
They obtained state-of-the-art results for many benchmarks. This model is the base of CodeT5 and is used for RQ2 as a general-LLM. 

\textbf{Llama2.} Llama2~\cite{llama2-touvron2023llama} is the second version of Llama \cite{Touvron2023LLaMAOA}, a series of foundational large language models proposed in 7 to 65 billion parameter budgets, which can outperform larger models such as GPT-3 \cite{gpt3-NEURIPS2020_1457c0d6} with 175 billion parameters and keep up with much larger models such as PaLM \cite{Chowdhery2022PaLMSL} with 540 billion parameters. Llama's architecture is inspired by various advancements from different models such as SwiGLUE \cite{shazeer2020glu} and Rotary Embeddings \cite{SU2024127063}. In contrast with the GPT series, the open-source nature of Llama models allows independent users and researchers to use and experiment with them more easily. Compared to Llama, Llama2 was trained on 40\% more training data and allows for a larger context length. 
Choosing Llama2 over Llama provides us with a fairer comparison in RQ2, given that CodeLlama is based on Llama2.

We employ T5 and the base version of CodeT5 with 220 million parameters and the 7 billion parameters Llama2 and CodeLlama-base in all of our experiments. 
It is worth mentioning that our experiments are mainly focused on the comparison of the performance of PEFT methods within the same base models, not across different based models with varied parameter budgets, hence, the difference in the number of parameters among the models is not a threat to the validity of our experiments.
Our implementation of the base models involves initializing model implementations similar to HuggingFace's for T5 and Llama2 by publicly available CodeT5 and CodeLlama checkpoints.

To reduce the effect of different prompting techniques on the performance of the instruction-tuned models for CodeLlama, we use the base version of Llama2 and CodeLlama models.
Additionally, we sample models in zero-shot settings without any prompt engineering. These settings eliminate unnecessary confounding variables, ensuring that the comparison of the results is fair among all PEFT methods.

\subsection{PEFT Methods} \label{sec:pefts}

In our experiments, we employ LoRA, Compacter and IA$^3$ in combination with the fixed-parameter models, CodeT5 and CodeLlama (and T5 and Llama2 for RQ2), across all experiments. 
We choose to incorporate LoRA in our experiments due to its stronger performance compared to other PEFT methods, as shown by previous researchers in various settings \cite{chen-etal-2022-revisiting, Zhuo2024astraios, 2023arXiv230810462W}. Given this, LoRA is a suitable representative of state-of-the-art PEFT methods for studying knowledge transfer.
Compacter and IA$^3$ on the other hand are relatively unexplored PEFT methods, specifically for software engineering tasks. By comparing them to LoRA, our study can provide valuable insight into their performance, specifically when utilized for transferring knowledge to unseen languages (RQ3).
We only select these three PEFT techniques due to restrictions of computational resources of fine-tuning LLMs, and since previous studies have shown these are SOTA approaches in terms of their achieved performance in the NLP field. As these three PEFT methods are under-explored in SE, we study them in our experiments.

As implementation of the PEFT methods is not within our scope, we rely on third-party implementations in this study. For LoRA and IA$^3$, we use HuggingFace's \emph{peft} library \cite{peft-library}, and for Compacter we use the \emph{AdapterHub} \cite{adapterhub-poth2023adapters} platform to facilitate composing all model configurations. Given the large number of experiments resulting from various combinations of models, programming languages, target tasks and PEFT methods, we refrain from experimenting with multiple hyperparameters available for each PEFT method and instead rely on the hyperparameters proposed by the authors of each PEFT method and third-party libraries as optimal defaults.

Moreover, we use the 16-bit \emph{brain floating point} data type to greatly reduce runtime and memory usage during fine-tuning, without impacting performance significantly. While LoRA and Compacter's implementations support this data type, IA$^3$'s implementation lacks support. As a fallback, we use traditional 32-bit floating point data types for IA$^3$. This does not impact the results obtained for IA$^3$, however, this is not the case for its runtime and memory usage. Thus, we only report the performance metrics of IA$^3$ and eliminate this PEFT when discussing the computational resource efficiency of the PEFT methods in the discussion section (Section \ref{sec:discussions}). 

\subsection{Statistical Test.} 

The statistical tests are used to determine if the differences in performance between PEFT-infused models and their baselines are significant.
To assess the significance of the results of this study's experiments, we report the statistical tests using ANOVA and pair-wise t-tests \cite{anova} for the results obtained from each PEFT and its respective baseline. We employ bootstrap sampling with replacement, $1000$ repetitions and sample sizes of $100$ and ensure the normality of distributions by the Shapiro–Wilk normality test \cite{shapiro-SHAPIRO1965}.
We limit the reports on the p-values of the BLEU-4. Specifically, we omit the statistical test reports on the HumanEval benchmarks for Python and R code generation tasks for the Pass@1 metric, as we use deterministic generation for these benchmarks and eliminate randomness.

\subsection{Experimental Setup}
We conduct the experiments using one Nvidia 40GB A100 GPU to fine-tune and test all model configurations in this study. We use HuggingFace's \emph{peft} library \cite{peft-library} and \emph{AdapterHub} \cite{adapterhub-poth2023adapters} platform to facilitate composing all model configurations.
Throughout the experimentation process, the Llama and T5 family of models undergo fine-tuning with learning rates of $2\mathrm{e}{-4}$ and $3\mathrm{e}{-3}$, respectively, for a total of five epochs for smaller datasets (i.e., SPP and R datasets) and one epoch for CodeSearchNet, with early stopping mechanisms in place to ensure optimal performance. We use $500$ warmup steps and a batch size of $16$ for the larger models (i.e., Llama2 and CodeLlama) and $100$ warmup steps and a batch size of $64$ for the smaller models (i.e., T5 and CodeT5). Additionally, we utilize 4-bit float quantized base models to facilitate fine-tuning.

Our approach aligns with default implementation, following the established guidelines for setting hyperparameters specific to these PEFT methods. Following previous studies \cite{weyssow2023exploring}, for LoRA and IA$^3$, we add matrices to the attention layers and maintain a dimension rank of $8$ and $\alpha = 16$ and for Compacter, we use a $phm$ dimension and rank of $4$ and $1$, respectively. For other hyperparameters, we rely on libraries' defaults.

For datasets that do not come in pre-defined train/validation/test splits (i.e., Python and R code generation datasets), we randomly select $4.95\%$ and $1\%$ for testing and validation, respectively. These proportions are necessary for effective training since the original datasets are relatively small (less than $40,000$ samples), especially for the larger Llama family of models.

\section{Results}\label{sec:results}
\newcolumntype{"}{@{\hskip\tabcolsep\vrule width 1pt\hskip\tabcolsep}}
\begin{table}
    \centering
    \setlength{\aboverulesep}{0pt}
    \setlength{\belowrulesep}{0pt}
    \resizebox{\textwidth}{!}{
        \begin{tabular}{l|c|c|c|c|c|c|c"c"|c}
            \toprule
            \multirow{2}*{\textbf{PEFT}} & \multirow{2}*{\pbox{2cm}{\textbf{Parameters}\\ \textbf{trained \%}}} & \multicolumn{7}{c"|}{\emph{\textbf{Seen}}} & \multicolumn{1}{c}{\emph{\textbf{Unseen}}} \\
            \cmidrule{3-10}
            & & \multirow{1}*{Python} & \multirow{1}*{Go} & \multirow{1}*{Java} & \multirow{1}*{Javascript} & \multirow{1}*{Ruby} & \multirow{1}*{PHP} & \multirow{1}*{\textbf{Average}} & \multirow{1}*{R} \\
            \hline 
            \multicolumn{10}{c}{\cellcolor{green!10}\emph{CodeT5}}\\
            \midrule
            Full &100\% &\cellcolor[HTML]{e69138}19.65 &\cellcolor[HTML]{e69138}19.10 &\cellcolor[HTML]{e69138}19.92 &\cellcolor[HTML]{e69138}15.48 &\cellcolor[HTML]{ffe69c}14.73 &\cellcolor[HTML]{e69138}25.14 &\cellcolor[HTML]{e69138}19.00 &\cellcolor[HTML]{f5c16f}6.72 \\
            \midrule
            Compacter &0.051\% &\cellcolor[HTML]{ffeaab}19.30 &\cellcolor[HTML]{ffeebc}18.36 &\cellcolor[HTML]{ffe9a9}19.20 &\cellcolor[HTML]{ffebaf}15.16 &\cellcolor[HTML]{e69138}15.05 &\cellcolor[HTML]{ffeaad}24.54 &\cellcolor[HTML]{ffe9a8}18.60 &\cellcolor[HTML]{ffefbf}5.30 \\
            LoRA &0.595\% &\cellcolor[HTML]{f2b865}19.55 &\cellcolor[HTML]{efb05b}18.94 &\cellcolor[HTML]{fad283}19.47 &\cellcolor[HTML]{e8963e}15.48 &\cellcolor[HTML]{fee094}14.76 &\cellcolor[HTML]{f9ce7e}24.80 &\cellcolor[HTML]{f5c372}18.83 &\cellcolor[HTML]{e69138}7.63 \\
            IA$^3$ &0.037\% &\cellcolor[HTML]{fff2cc}19.06 &\cellcolor[HTML]{fff2cc}18.22 &\cellcolor[HTML]{fff2cc}18.88 &\cellcolor[HTML]{fff2cc}14.94 &\cellcolor[HTML]{fff2cc}14.34 &\cellcolor[HTML]{fff2cc}24.32 &\cellcolor[HTML]{fff2cc}18.29 &\cellcolor[HTML]{fff2cc}5.04 \\
            \midrule
            
            \multicolumn{10}{c}{\cellcolor{green!10}\emph{T5}}\\
            \midrule
            Full &100\% &\cellcolor[HTML]{e69138}18.48 &\cellcolor[HTML]{e69138}18.13 &\cellcolor[HTML]{e69138}17.57 &\cellcolor[HTML]{e69138}13.80 &\cellcolor[HTML]{eda651}12.58 &\cellcolor[HTML]{e69138}22.66 &\cellcolor[HTML]{e69138}17.20 &\cellcolor[HTML]{e69138}13.30 \\
            \midrule
            Compacter &0.051\% &\cellcolor[HTML]{fff2cc}16.61 &\cellcolor[HTML]{fff2cc}15.41 &\cellcolor[HTML]{fff0c4}15.50 &\cellcolor[HTML]{ffeebc}11.58 &\cellcolor[HTML]{ffeaac}11.81 &\cellcolor[HTML]{fff2cc}20.73 &\cellcolor[HTML]{fff0c2}15.27 &\cellcolor[HTML]{fff0c2}3.35 \\
            LoRA &0.595\% &\cellcolor[HTML]{efaf5a}18.10 &\cellcolor[HTML]{f3bb69}17.25 &\cellcolor[HTML]{f8cc7c}16.46 &\cellcolor[HTML]{f7c877}12.75 &\cellcolor[HTML]{e69138}12.70 &\cellcolor[HTML]{f6c574}21.82 &\cellcolor[HTML]{f4be6c}16.51 &\cellcolor[HTML]{fee093}4.65 \\
            IA$^3$ &0.037\% &\cellcolor[HTML]{fff1c8}16.68 &\cellcolor[HTML]{fff2c9}15.47 &\cellcolor[HTML]{fff2cc}15.40 &\cellcolor[HTML]{fff2cc}11.29 &\cellcolor[HTML]{fff2cc}11.14 &\cellcolor[HTML]{fff2cb}20.75 &\cellcolor[HTML]{fff2cc}15.12 &\cellcolor[HTML]{fff2cc}3.19 \\
            \midrule
            \multicolumn{10}{c}{\cellcolor{blue!10}\emph{CodeLlama}}\\
            \midrule
            Zero-Shot &0.000\% &\cellcolor[HTML]{fff2cc}0.79 &\cellcolor[HTML]{fff2cc}1.01 &\cellcolor[HTML]{fff2cc}1.00 &\cellcolor[HTML]{fff2cc}0.82 &\cellcolor[HTML]{fff2cc}0.97 &\cellcolor[HTML]{fff2cc}0.58 &\cellcolor[HTML]{fff2cc}0.86 &\cellcolor[HTML]{fff2cc}1.10 \\
            \midrule
            Compacter &0.012\% &\cellcolor[HTML]{fbd688}20.21 &\cellcolor[HTML]{fdde91}19.86 &\cellcolor[HTML]{f9d080}19.89 &\cellcolor[HTML]{f7ca7a}15.28 &\cellcolor[HTML]{f1b360}15.70 &\cellcolor[HTML]{f9d181}25.09 &\cellcolor[HTML]{f8cb7b}19.34 &\cellcolor[HTML]{ffe498}5.36 \\
            LoRA &0.093\% &\cellcolor[HTML]{e69138}20.86 &\cellcolor[HTML]{e69138}20.97 &\cellcolor[HTML]{e69138}20.77 &\cellcolor[HTML]{e69138}16.45 &\cellcolor[HTML]{e69138}16.47 &\cellcolor[HTML]{e69138}26.06 &\cellcolor[HTML]{e69138}20.26 &\cellcolor[HTML]{e69138}10.62 \\
            IA$^3$ &0.006\% &\cellcolor[HTML]{ffe69a}19.91 &\cellcolor[HTML]{ffe69a}19.65 &\cellcolor[HTML]{ffe69a}19.26 &\cellcolor[HTML]{ffe69c}14.16 &\cellcolor[HTML]{ffe79e}13.43 &\cellcolor[HTML]{ffe69a}24.45 &\cellcolor[HTML]{ffe69b}18.48 &\cellcolor[HTML]{ffe69b}5.17 \\
            \midrule
            \multicolumn{10}{c}{\cellcolor{blue!10}\emph{Llama2}}\\
            \midrule
            Zero-Shot &0.000\% &\cellcolor[HTML]{fff2cc}0.91 &\cellcolor[HTML]{fff2cc}1.06 &\cellcolor[HTML]{fff2cc}0.93 &\cellcolor[HTML]{fff2cc}0.76 &\cellcolor[HTML]{fff2cc}0.87 &\cellcolor[HTML]{fff2cc}0.69 &\cellcolor[HTML]{fff2cc}0.87 &\cellcolor[HTML]{fff2cc}1.04 \\
            \midrule
            Compacter &0.012\% &\cellcolor[HTML]{f2b865}18.75 &\cellcolor[HTML]{fcd88a}18.53 &\cellcolor[HTML]{e7943b}18.71 &\cellcolor[HTML]{f9d181}14.08 &\cellcolor[HTML]{eeab56}14.51 &\cellcolor[HTML]{fad283}23.18 &\cellcolor[HTML]{f4bd6b}17.96 &\cellcolor[HTML]{fee295}5.19 \\
            LoRA &0.093\% &\cellcolor[HTML]{e69138}19.06 &\cellcolor[HTML]{e69138}19.15 &\cellcolor[HTML]{e69138}18.72 &\cellcolor[HTML]{e69138}14.86 &\cellcolor[HTML]{e69138}15.21 &\cellcolor[HTML]{e69138}24.10 &\cellcolor[HTML]{e69138}18.52 &\cellcolor[HTML]{e69138}9.52 \\
            IA$^3$ &0.006\% &\cellcolor[HTML]{ffe69b}18.03 &\cellcolor[HTML]{ffe69a}18.30 &\cellcolor[HTML]{ffe69b}17.87 &\cellcolor[HTML]{ffe69b}13.56 &\cellcolor[HTML]{ffe7a0}11.26 &\cellcolor[HTML]{ffe69a}22.63 &\cellcolor[HTML]{ffe69b}16.94 &\cellcolor[HTML]{ffe69c}4.80 \\
            \bottomrule
        \end{tabular}
    }
    \caption{The BLEU-4 scores and percentage of trained parameters of the models fine-tuned with Compacter, LoRA and IA$^3$ for code summarization per programming language and their average across the six seen programming languages. T5 and CodeT5 are fully fine-tuned and Llama2 and CodeLlama are sampled in a zero-shot setting as respective baselines. "\emph{Seen}" denotes languages included in the pre-training phase of corresponding base models. "\emph{Unseen}" denotes languages excluded from the pre-training phase of corresponding base models. A higher score is better. The darker the cell color, the better the score; heatmap is presented per each LLM and programming language.}
    
    \label{tab:rq1-sum}
\end{table}

This section explains the results. 
For better interpretation and comparison, we discuss the results of RQ1 and RQ2 separated by the tasks and model architectures together. We will first describe the obtained results for code summarization for T5 family of models and then for the Llama family of models in section~\ref{sec:rq1-2-summarization}. A similar process is followed to explain code generation results for both RQ1 and RQ2 in section~\ref{sec:rq1-2-generation}. 
Then, we discuss the RQ3 results for knowledge transfer to unseen language, R, for both code summarization and code generation, in section~\ref{sec:rq3-r}. 

For all experiments in which we report BLEU scores, including code summarization and code generation, we applied statistical tests to determine whether the obtained results were significant. For all cases, we found that the performance differences in terms of the BLEU-4 scores between each PEFT method and its respective baselines are significant, with $P < 0.001$.
As the significant differences are a common observation, we report the p-value here and in the following, we discuss the results for each RQ. 
Please note that as the Pass@1 results are deterministic, there is no need to apply statistical tests and comparing the Pass@1 scores suffice. 

\subsection{Code Summarization} \label{sec:rq1-2-summarization}

\textbf{T5 Family of Models.} 
Table~\ref{tab:rq1-sum} illustrates the performance of each PEFT technique for the code summarization task. Among the PEFT techniques applied to CodeT5, an encoder-decoder code-LLM, LoRA achieves the highest average BLEU-4 score among the three PEFT techniques across all six programming languages, with an average score of $18.83$, followed by Compacter with $18.60$ and IA$^3$ with $18.29$. Notably, the fully fine-tuned CodeT5 outperforms all three PEFT techniques in this setting; Compared to full fine-tuning, LoRA, Compacter and IA$^3$ degrade summarization performance by $0.89\%$, $2.11\%$ and $3.74\%$, respectively. 
The observed order of performance obtained by the three PEFT techniques is directly correlated with the percentage of trained parameters per each PEFT technique, where LoRA with the largest trainable parameter budget ($0.595\%$ of total parameters) performs best, Compacter with the second largest parameter budget ($0.051\%$ of total parameters) stands in the middle performance-wise, and IA$^3$ with the smallest trainable parameter budget ($0.037\%$ of total parameters) perform worst. 

\begin{figure}[ht]
    \centering
    \includegraphics[width=\textwidth]{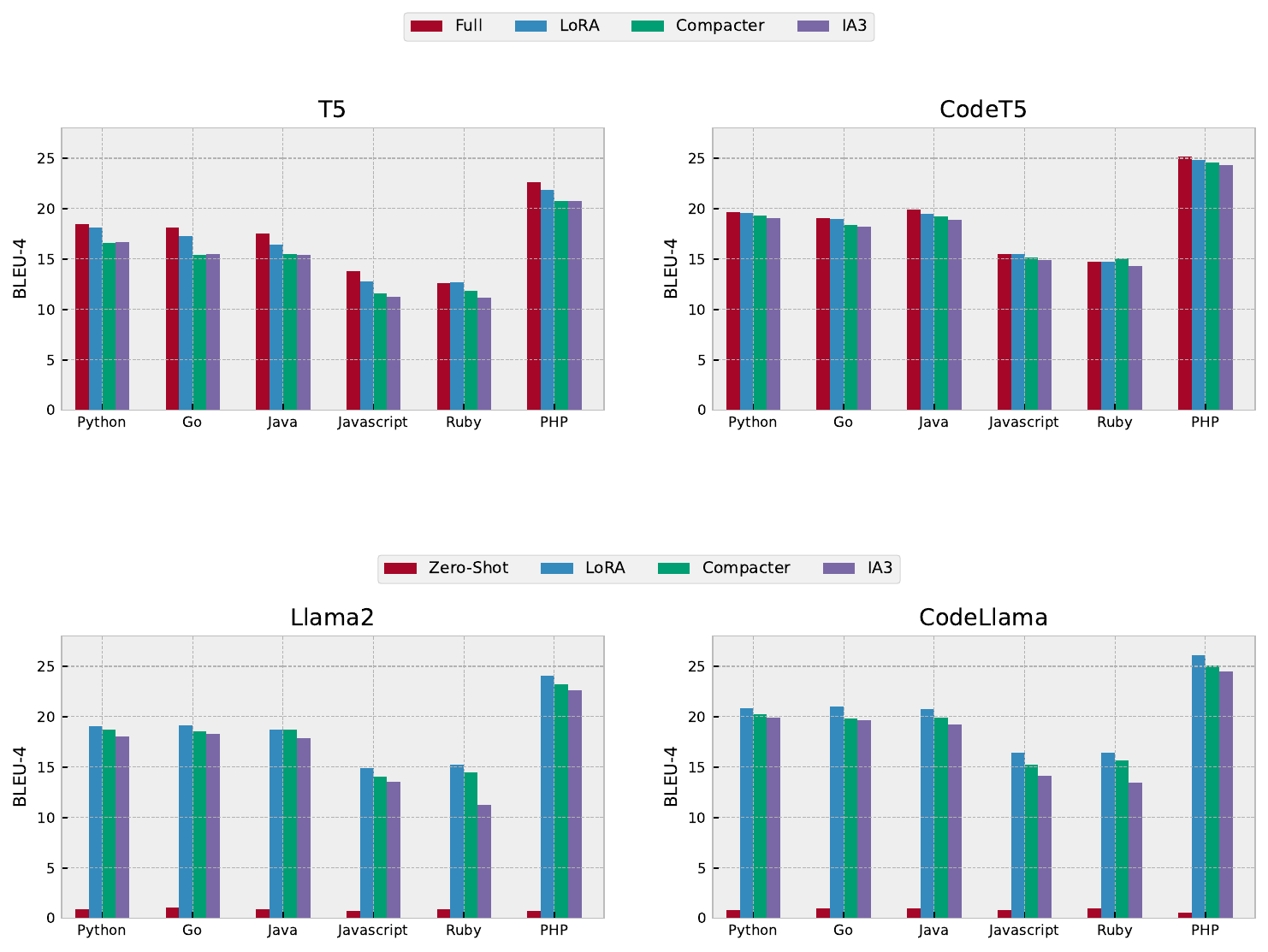}
    \caption{BLEU-4 scores of LoRA, Compacter and IA$^3$ per programming language for code summarization on T5 (top left), CodeT5 (top right), Llama2 (bottom left) and CodeLlama (bottom right) and their respective baselines.}
    \label{fig:sum-compare}
\end{figure}

The performance trend of all three PEFT techniques remains consistent across all programming languages for code summarization on CodeT5, with an exception for Ruby, where Compacter outperforms other fine-tuning techniques with a BLEU-4 score of $15.05$ and Javascript, where LoRA performs on par with full fine-tuning, obtaining a BLEU-4 score of $15.48$ (see top right plot of Fig. \ref{fig:sum-compare}).
For the T5 model, LoRA with an average BLEU-4 score of $16.51$ outperforms Compacter with an average score of $15.27$ and IA$^3$ with an average score of $15.12$ on T5, resulting in a similar PEFT technique effectiveness ranking when compared to CodeT5, while full fine-tuning T5 yields the highest BLEU-4 score of $17.20$, outperforming LoRA, Compacter and IA$^3$ for code summarization by $4.18\%$, $12.64\%$ and $13.76\%$, respectively.

While among all the PEFT techniques applied to T5, LoRA is dominant for all six programming languages, Compacter and IA$^3$ do not perform consistently relative to each other (Fig. \ref{fig:sum-compare}, top left).
Specifically, Compacter is more advantageous in Java, Javascript and Ruby, while for other programming languages, IA$^3$ slightly outperforms Compacter. Overall, however, both Compacter and IA$^3$ perform very closely in terms of BLEU-4 score, with no clear top performer on the T5 model.

For code summarization, all three PEFT techniques prove more beneficial on CodeT5 than T5 when compared to their fully fine-tuned counterparts. On CodeT5, the most effective PEFT technique (i.e. LoRA) results in a $0.89\%$ performance drop when compared to full fine-tuning, while on T5, this performance gap increases to $4.01\%$. This difference in effectiveness can be rooted in the fact that PEFT techniques utilize a limited budget of trainable parameters during fine-tuning. Due to this limitation, they are more suitable for leveraging present knowledge in the base model and adapting it for better usage of that knowledge than teaching the model new information.

\textbf{Llama Family of Models.}
Table~\ref{tab:rq1-sum} presents the results for each PEFT technique applied to CodeLlama and Llama2 in the code summarization task. For CodeLlama, LoRA achieves the highest average BLEU-4 score across all six programming languages, with an average score of $20.26$, followed by Compacter at $19.34$ and IA$^3$ at $18.48$. This mirrors the pattern seen in the T5 family models, where LoRA, with its relatively larger percentage of trainable parameters ($0.093\%$), outperforms the other PEFT techniques. In comparison to zero-shot performance, where CodeLlama scores an average of just $0.86$, all PEFT techniques significantly boost the BLEU-4 scores, emphasizing the importance of fine-tuning.
In terms of individual language performance on CodeLlama, LoRA demonstrates superiority across the board, performing best in all languages, particularly in Javascript, where LoRA surpasses Compacter and IA$^3$ by $7.66\%$ and $16.17\%$, respectively (Fig. \ref{fig:sum-compare}, bottom right). 

For Llama2, the overall ranking of PEFT techniques remains consistent with CodeLlama. LoRA achieves the highest average score of $18.52$, followed by Compacter at $17.96$ and IA$^3$ at $16.94$. As with CodeLlama, zero-shot results for Llama2 are substantially lower, with an average score of just $0.87$. However, despite the highly comparable performance of CodeLlama and Llama2 in the zero-shot setting in terms of BLEU-4, the performance gains for PEFT techniques on Llama2 are slightly lower than those observed for CodeLlama. This suggests that CodeLlama, specifically designed for coding tasks, benefits more from these fine-tuning techniques than the general-purpose Llama2.

LoRA maintains its dominance when applied to Llama2 where it consistently performs better than Compacter and IA$^3$. Unlike when applied to CodeLlama however, Compacter better closes the performance gap with LoRA on Llama2, specifically for Python and Java, where it lags behind only $1.65\%$ and $0.05\%$, respectively (Fig. \ref{fig:sum-compare}, bottom left).

IA$^3$, despite its small parameter budget ($0.006\%$), manages to keep pace in some cases but generally lags behind LoRA and Compacter across most languages.
While IA$^3$ performs close to Compacter on the T5 and CodeT5 models, the performance gap is slightly larger on the Llama family of models, which might be related to its small parameter budget.

Overall, LoRA has the best score across a broad range of programming languages, which might be due to its larger trainable parameter allocation compared to other PEFT techniques.
In Section \ref{sec:discussions}, we provide an analysis in a specific setting to compare the performance of LoRA and Compacter with respect to their parameter budget. 

\begin{table}
    \centering
    \setlength{\aboverulesep}{0pt}
    \setlength{\belowrulesep}{0pt}
        \begin{tabular}{l|c|c|c|c|c}
            \toprule
            \multirow{2}*{\textbf{PEFT}} & \multirow{2}*{\pbox{3cm}{\textbf{Parameters}\\ \textbf{trained \%}}} & \multirow{2}*{\textbf{CodeBLEU}} & \multirow{2}*{\textbf{BLEU-4}} & \multirow{2}*{\textbf{EM@1}} & \multirow{2}*{\textbf{Pass@1}} \\
            &&&&&\\
            \hline
            \multicolumn{6}{c}{\cellcolor{green!10}\emph{CodeT5}}\\

            \midrule
            Full &100\% &\cellcolor[HTML]{e69138}25.60 &\cellcolor[HTML]{e7943b}41.69 &\cellcolor[HTML]{e89840}14.56 &\cellcolor[HTML]{e69138}6.10 \\
            \midrule
            Compacter &0.051\% &\cellcolor[HTML]{fff2cc}6.05 &\cellcolor[HTML]{ffe498}29.23 &\cellcolor[HTML]{fddd90}10.79 &\cellcolor[HTML]{fee296}2.44 \\
            LoRA &0.595\% &\cellcolor[HTML]{eba14b}22.32 &\cellcolor[HTML]{e69138}42.01 &\cellcolor[HTML]{e69138}14.90 &\cellcolor[HTML]{ea9f48}5.49 \\
            IA$^3$ &0.037\% &\cellcolor[HTML]{ffe296}8.83 &\cellcolor[HTML]{fff2cc}28.20 &\cellcolor[HTML]{fff2cc}8.97 &\cellcolor[HTML]{fff2cc}1.83 \\
            \midrule
            \multicolumn{6}{c}{\cellcolor{green!10}\emph{T5}}\\
            \midrule
            Full &100\% &\cellcolor[HTML]{e69138}22.59 &\cellcolor[HTML]{e69138}42.74 &\cellcolor[HTML]{fff2cc}0.00 &\cellcolor[HTML]{fff2cc}0.00 \\
            \midrule
            Compacter &0.051\% &\cellcolor[HTML]{ffe599}4.19 &\cellcolor[HTML]{ffe599}15.93 &\cellcolor[HTML]{fff2cc}0.00 &\cellcolor[HTML]{fff2cc}0.00 \\
            LoRA &0.595\% &\cellcolor[HTML]{f4be6c}12.76 &\cellcolor[HTML]{efad58}34.01 &\cellcolor[HTML]{fff2cc}0.00 &\cellcolor[HTML]{fff2cc}0.00 \\
            IA$^3$ &0.037\% &\cellcolor[HTML]{fff2cc}3.69 &\cellcolor[HTML]{fff2cc}15.36 &\cellcolor[HTML]{fff2cc}0.00 &\cellcolor[HTML]{fff2cc}0.00 \\
            \midrule
            \multicolumn{6}{c}{\cellcolor{blue!10}\emph{CodeLlama}}\\
            \midrule
            Zero-Shot &0.000\% &\cellcolor[HTML]{ffe397}26.37 &\cellcolor[HTML]{fff2cc}5.28 &\cellcolor[HTML]{fff2cc}0.00 &\cellcolor[HTML]{fff2cc}4.88 \\
            \midrule
            Compacter &0.012\% &\cellcolor[HTML]{fff2cc}26.17 &\cellcolor[HTML]{f3bd6b}43.31 &\cellcolor[HTML]{f5c372}14.63 &\cellcolor[HTML]{eda751}21.95 \\
            LoRA &0.093\% &\cellcolor[HTML]{e69138}31.10 &\cellcolor[HTML]{e69138}47.67 &\cellcolor[HTML]{e69138}16.72 &\cellcolor[HTML]{e69138}26.83 \\
            IA$^3$ &0.006\% &\cellcolor[HTML]{ffe599}26.25 &\cellcolor[HTML]{f4c06e}43.07 &\cellcolor[HTML]{f3bb68}14.97 &\cellcolor[HTML]{ffe498}7.93 \\
            \midrule
            \multicolumn{6}{c}{\cellcolor{blue!10}\emph{Llama2}}\\
            \midrule
            Zero-Shot &0.000\% &\cellcolor[HTML]{ffe396}23.67 &\cellcolor[HTML]{fff2cc}4.30 &\cellcolor[HTML]{fff2cc}0.00 &\cellcolor[HTML]{fff2cc}2.44 \\
            \midrule
            Compacter &0.012\% &\cellcolor[HTML]{f4bf6d}26.02 &\cellcolor[HTML]{f0b35f}42.25 &\cellcolor[HTML]{eeac57}14.23 &\cellcolor[HTML]{f0b05b}12.80 \\
            LoRA &0.093\% &\cellcolor[HTML]{e69138}29.07 &\cellcolor[HTML]{e69138}46.11 &\cellcolor[HTML]{e69138}15.71 &\cellcolor[HTML]{e69138}15.24 \\
            IA$^3$ &0.006\% &\cellcolor[HTML]{fff2cc}21.86 &\cellcolor[HTML]{f6c776}39.95 &\cellcolor[HTML]{f9d080}12.20 &\cellcolor[HTML]{fddd90}9.15 \\
            \bottomrule
        \end{tabular}
    \caption{The CodeBLEU, BLEU-4, EM@1 and Pass@1 scores and percentage of trained parameters of the models fine-tuned with Compacter, LoRA and IA$^3$ for code generation in Python. T5 and CodeT5 are fully fine-tuned and Llama2 and CodeLlama are sampled in a zero-shot setting as respective baselines. Higher is better. The darker the color, the better the score; heatmap is presented per each model and metric.}
    \label{tab:rq1-2-gen}
\end{table}

\begin{tcolorbox}[customsummarybox]
For the code summarization in the six seen languages, LoRA performs best, followed by Compacter which slightly outperforms IA$^3$ on CodeT5 and T5.

A similar performance trend appears on CodeLlama and Llama2, though the gap between Compacter and IA$^3$ is larger for the Llama models.

All three PEFT methods demonstrate better effectiveness on code-LLMs, but the performance gap between the two categories of models suggests that all three methods are effective in code summarization knowledge transfer on general-LLMs for the seen languages.
\end{tcolorbox}

\subsection{Code Generation} \label{sec:rq1-2-generation}

The results of all models for code generation are shown in Table~\ref{tab:rq1-2-gen}.

In addition to BLEU-4 which was also used for evaluating code summarization, we utilize CodeBLEU, EM@1 and Pass@1 for code generation. However, we consider Pass@1 as the main performance metric for code generation given that HumanEval evaluation problems are designed to represent real-world scenarios not included in the training datasets.
Moreover, we refrain from discussing EM@1 in detail, as some EM@1 scores are zero, but most EM@1 results are generally tightly coupled with BLEU-4 as the main textual similarity metric.

\textbf{T5 Family of Models.}
The gap in performance between LoRA, Compacter and IA$^3$ increases for Python code generation in contrast to code summarization, as illustrated in Table \ref{tab:rq1-2-gen}. 

On CodeT5, besides fully fine-tuning, which proves most effect with a Pass@1 score of $6.10$, LoRA is the best-performing PEFT followed by Compacter and IA$^3$ with Pass@1 scores of $5.49$, $2.44$ and $1.83$, respectively. 

Interestingly, when strictly looking at the BLEU-4 and EM@1 scores, LoRA appears to outperform all other fine-tuning techniques with a BLEU-4 score of $42.01$, followed by fully fine-tuning with a score of $41.69$, Compacter with a score of $29.23$ and IA$^3$ with a score of $28.20$. The EM@1 scores follow an identical trend.
While the higher BLEU-4 and EM@1 scores of LoRA compared to the fully fine-tuned CodeT5 suggests that LoRA performs better in generating textually similar to ground truth samples, it does not mean it generalizes better to out-of-distribution code generation problems, judged by higher Pass@1 score for the fully fine-tuned model.
LoRA is the next high Pass@1, with a score of $5.49$ compared to $6.10$ for the fully fine-tuned CodeT5.

Moreover, it can be observed that despite relatively close BLEU-4 scores of Compacter and IA$^3$ ($29.23$ and $28.20$, respectively), their CodeBLEU scores are more apart, with scores of $6.05$ and $8.83$, respectively. This discrepancy in difference and trend can be attributed to the fact that the BLEU-4 score is strictly concerned with textual similarity. On the other hand, besides textual similarity, CodeBLEU additionally accounts for the type of tokens and the similarity in the abstract tree and the data flow. This means that between Compacter and IA$^3$, while the former generates more textually similar to ground truth code snippets, the latter generations are closer to the ground truth in terms of functionality. Consequently, the same is true for LoRA and fully fine-tuning with CodeBLEU scores of $22.32$ and $25.60$ respectively, where LoRA's generations are more similar textually and less similar functionally to the ground truth compared to fully fine-tuning.

T5's performance in code generation is notably poor, as none of the code snippets produced by any of the fine-tuning methods are capable of successfully solving any HumanEval problems, all with Pass@1 equal to zero. Nevertheless, besides fully fine-tuning with a BLEU-4 score of $42.74$ and a CodeBLEU score of $22.59$, LoRA is the best-performing fine-tuning technique with a BLEU-4 score of $34.01$ and a CodeBLEU score of $12.76$. Among Compacter and IA$^3$, the former's generations are slightly more aligned with the ground truth in terms of textual similarity, as observed by a slightly better BLEU-4 score, while the latter's generations are more functionally similar to the ground truth, as indicated by higher CodeBLEU.

The zero Pass@1 scores obtained by all fine-tuned T5 models suggest that even though these models learn to generate code snippets from their training dataset's distribution, none of the fine-tuning techniques are capable of adapting and generalizing to real-world code generation problems. LoRA on CodeT5 is, however, capable of generalizing to real-world problems to some extent, reaching close to fully fine-tuned CodeT5 on the HumanEval benchmark.

Overall, for CodeT5, all fine-tuning techniques obtain very low Pass@1 scores; indicating that T5 family are not capable of generating executable code.

\textbf{Llama Family of Models.}
Aligned with previous observations, CodeLlama is the best-performing model for Python code generation when fine-tuned using LoRA. LoRA yields the highest HumanEval Pass@1 score of $26.83$, followed by $21.95$ of Compacter and $7.93$ of IA$^3$, which are significant performance boosts compared to the baseline zero-shot setting with a score of $4.88$.

Considering other metrics, LoRA achieves the highest BLEU-4 score of $47.67$, CodeBLEU score of $31.10$ and EM@1 score of $16.72$. Similar to T5 family models, Compacter and IA$^3$ are ordered differently when ranked by BLEU-4 CodeBLEU, even though the discrepancy is not as large. Compacter slightly outperforms IA$^3$ in textual similarity (i.e., BLEU-4), while IA$^3$ slightly outperforms Compacter in functionality similarity (i.e., CodeBLEU). 
In comparison, the differences in textual and functional similarities of CodeLlama's zero-shot setting generations are more noticeable, with a BLEU-4 score of $5.28$ and a CodeBLEU score of $26.37$.

The general decoder-only Llama2's code generation results are similar to CodeLlama in terms of comparison between the PEFT metrics. On Llama2, LoRA gains the highest Pass@1 of $15.24$, followed by Compacter with $12.80$ and IA$^3$ with $9.15$. The zero-shot scores of the Llama2 model obtain a relatively low Pass@1 score of $2.44$.
The BLEU-4, CodeBLEU and EM@1 scores of PEFT methods are ranked in the same order, 
LoRA being the highest, followed by Compacter and IA$^3$ for all three metrics. 
Similar to CodeLlama, Llama2 obtains a much higher CodeBLEU score compared to its BLEU-4 score in the zero-shot setting.

For both Llama2 and CodeLlama in the zero-shot setting, we observe a discrepancy between the BLEU-4 and CodeBLEU scores. This is due to the fact that before any fine-tuning, both models will not stop their generation after completing the target function's body, repeating the function definition until they hit the maximum token ceiling. Unlike BLEU-4 which penalizes predictions that are longer than the ground truth, CodeBLEU ignores the prediction length. This means that a significantly lengthy prediction can obtain a high CodeBLEU score along with a low BLEU-4 score, given it contains a large number of similar-to-reference code constructs.

Comparing the similar BLEU-4 and CodeBLEU scores of Llama2 and CodeLlama suggests all PEFT techniques are capable of transferring most of the information contained in the training dataset to adapt both models for its distribution adequately. On the other hand, when dealing with out-of-distribution real-world HumanEval problems, CodeLlama still outperforms Llama2 by a large margin as obvious by Pass@1 scores. 

\begin{tcolorbox}[customsummarybox, float=h]
For code generation in the six seen languages, LoRA significantly outperforms Compacter and IA$^3$ on the T5 and Llama family of models. While Compacter and IA$^3$ produce similar results in some cases, the former generally achieves higher functional accuracy (i.e., the Pass@1 score).

All three PEFT methods consistently show better effectiveness when applied to code-LLMs, with an exception for IA$^3$, which performs slightly better in terms of functional accuracy (i.e., the Pass@1 score) when applied to Llama2 compared to CodeLlama.
\end{tcolorbox}

\subsection{Knowledge Transfer to Unseen Language R} \label{sec:rq3-r}

R serves as an unseen programming language for evaluating each PEFT technique's ability to transfer knowledge to unseen domains. Following the previous structure, we discuss R code summarization and R code generation separately, with results provided in Table \ref{tab:rq1-sum} and Table~\ref{tab:rq3-gen}, respectively. 

\subsubsection{R Code Summarization}

\textbf{T5 Family of Models.}
As shown in Table \ref{tab:rq1-sum}, for CodeT5, LoRA stands out as the top-performing technique, achieving a BLEU-4 score of $7.63$, which surpasses even the fully fine-tuned model’s score of $6.72$. Compacter and IA$^3$, though also effective, result in lower BLEU-4 scores of $5.30$ and $5.04$, respectively, demonstrating a noticeable drop in performance compared to LoRA and full fine-tuning.
For the T5 model, however, a different trend is observed. Fully fine-tuning the T5 model results in the highest BLEU-4 score of $13.30$, significantly outperforming all PEFT techniques. Among the PEFT techniques, LoRA again shows the best performance with a BLEU-4 score of $4.65$, followed by Compacter at $3.35$ and IA$^3$ at $3.19$.

Notably, fully fine-tuning the model results in inconsistent performance between CodeT5 and T5. The fully fine-tuned CodeT5 performance is within the performance range of PEFT techniques on this model. Specifically, LoRA outperforms the fully fine-tuned CodeT5, despite updating a fraction of the model's parameters. In Contrast, the fully fine-tuned T5 model significantly outperforms the best-performing PEFT technique on this model, even surpassing the fully fine-tuned CodeT5.

Without considering the fully fine-tuned models, however, the performance ranking on the T5 family of models is consistent with the performance obtained for the six seen programming languages, with LoRA performing the best, followed by Compacter and IA$^3$.

\textbf{Llama Family of Models.}
When shifting focus to CodeLlama, the performance of PEFT techniques improves significantly compared to their zero-shot baseline. Without any fine-tuning, CodeLlama achieves a BLEU-4 score of $1.10$, but after applying LoRA, this score jumps to $10.62$, a substantial improvement that highlights LoRA's effectiveness for this model. Compacter and IA$^3$ also boost performance, with BLEU-4 scores of $5.36$ and $5.17$, respectively. However, LoRA clearly dominates in this setting, showing a much higher capability in enhancing CodeLlama's R summarization abilities compared to the other PEFT techniques.
Llama2 follows the same trend as CodeLlama, where the zero-shot BLEU-4 score is relatively low at $1.04$. Once again, LoRA emerges as the best-performing fine-tuning method with a BLEU-4 score of $9.52$, significantly improving the model’s ability to summarize R code. Compacter and IA$^3$, with scores of $5.19$ and $4.80$, respectively, also show improvement but fall short of LoRA’s performance. 

On the Llama family of models, the performance ranking of the PEFT techniques remains the same, where LoRA showcases the highest performance, followed by Compacter and IA$^3$. However, the performance gap between LoRA and the other two PEFT techniques on the Llama family of models increases when compared to the T5 family of models. 

Across all four models, LoRA consistently performs the best among the PEFT techniques for R code summarization. While full fine-tuning remains the most effective approach for T5, LoRA demonstrates the ability to achieve comparable results with fewer parameter updates. On CodeT5, LoRA even manages to outperform full fine-tuning, showcasing its potential to not only match but sometimes exceed traditional methods. Compacter and IA$^3$, while still beneficial, generally lag behind LoRA in terms of performance, particularly on the Llama family of models.

\begin{tcolorbox}[customsummarybox]
Compared to code summarization in the six seen languages, LoRA outperforms Compacter and IA$^3$ in the unseen setting, R code summarization, with a larger gap on both the T5 and Llama family of models. Although Compacter performs better than IA$^3$ in this setting, the performance gap between them is smaller compared to their gap with LoRA.

On the Llama family of models, all three PEFT methods achieve similar performance between the code-LLM and general-LLM variants. This performance difference increases for all three methods on the T5 family of models.
\end{tcolorbox}

\subsubsection{R Code Generation}
\begin{table}
    \centering
    \setlength{\aboverulesep}{0pt}
    \setlength{\belowrulesep}{0pt}
        \begin{tabular}{l|c|c|c|c}
            \toprule
            \multirow{2}*{\textbf{PEFT}} & \multirow{2}*{\pbox{3cm}{\textbf{Parameters}\\ \textbf{trained \%}}} & \multirow{2}*{\textbf{BLEU-4}} & \multirow{2}*{\textbf{EM@1}} & \multirow{2}*{\textbf{Pass@1}} \\
            &&&&\\
            \hline
            \multicolumn{5}{c}{\cellcolor{green!10}\emph{CodeT5}}\\
            \midrule
            Full &100\% &\cellcolor[HTML]{e69138}14.20 &\cellcolor[HTML]{e69138}0.27 &\cellcolor[HTML]{e69138}2.48 \\
            \midrule
            Compacter &0.051\% &\cellcolor[HTML]{fff2cc}8.70 &\cellcolor[HTML]{fff2cc}0.00 &\cellcolor[HTML]{fff2cc}0.62 \\
            LoRA &0.595\% &\cellcolor[HTML]{eca24c}13.15 &\cellcolor[HTML]{e69138}0.27 &\cellcolor[HTML]{efae59}1.86 \\
            IA$^3$ &0.037\% &\cellcolor[HTML]{ffe498}9.02 &\cellcolor[HTML]{fff2cc}0.00 &\cellcolor[HTML]{fff2cc}0.62 \\
            \midrule
            \multicolumn{5}{c}{\cellcolor{green!10}\emph{T5}}\\
            \midrule
            Full &100\% &\cellcolor[HTML]{e69138}12.74 &\cellcolor[HTML]{fff2cc}0.00 &\cellcolor[HTML]{fff2cc}0.00 \\
            \midrule
            Compacter &0.051\% &\cellcolor[HTML]{fff2cc}5.12 &\cellcolor[HTML]{fff2cc}0.00 &\cellcolor[HTML]{fff2cc}0.00 \\
            LoRA &0.595\% &\cellcolor[HTML]{f8ce7e}7.44 &\cellcolor[HTML]{fff2cc}0.00 &\cellcolor[HTML]{fff2cc}0.00 \\
            IA$^3$ &0.037\% &\cellcolor[HTML]{ffe598}5.45 &\cellcolor[HTML]{fff2cc}0.00 &\cellcolor[HTML]{fff2cc}0.00 \\
            \midrule
            \multicolumn{5}{c}{\cellcolor{blue!10}\emph{CodeLlama}}\\
            \midrule
            Zero-Shot &0.000\% &\cellcolor[HTML]{fff2cc}8.22 &\cellcolor[HTML]{fff2cc}0.00 &\cellcolor[HTML]{fff2cc}11.18 \\
            \midrule
            Compacter &0.012\% &\cellcolor[HTML]{f1b562}24.81 &\cellcolor[HTML]{e69138}0.91 &\cellcolor[HTML]{f7c979}18.01 \\
            LoRA &0.093\% &\cellcolor[HTML]{e69138}26.30 &\cellcolor[HTML]{f5c270}0.81 &\cellcolor[HTML]{e69138}20.50 \\
            IA$^3$ &0.006\% &\cellcolor[HTML]{f4bf6d}24.42 &\cellcolor[HTML]{e69138}0.91 &\cellcolor[HTML]{fbd789}17.39 \\
            \midrule
            \multicolumn{5}{c}{\cellcolor{blue!10}\emph{Llama2}}\\
            \midrule
            Zero-Shot &0.000\% &\cellcolor[HTML]{fff2cc}6.12 &\cellcolor[HTML]{fff2cc}0.00 &\cellcolor[HTML]{fff2cc}1.24 \\
            \midrule
            Compacter &0.012\% &\cellcolor[HTML]{f2b966}19.94 &\cellcolor[HTML]{f6c473}0.59 &\cellcolor[HTML]{f3bb69}8.70 \\
            LoRA &0.093\% &\cellcolor[HTML]{e69138}21.44 &\cellcolor[HTML]{e69138}0.75 &\cellcolor[HTML]{e69138}10.56 \\
            IA$^3$ &0.006\% &\cellcolor[HTML]{f5c270}19.60 &\cellcolor[HTML]{fbd586}0.54 &\cellcolor[HTML]{fbd789}7.45 \\
            \bottomrule
        \end{tabular}
    \caption{The CodeBLEU, BLEU-4, EM@1 and Pass@1 scores and percentage of trained parameters of the models fine-tuned with Compacter, LoRA and IA$^3$ for code generation in R. T5 and CodeT5 are fully fine-tuned and Llama2 and CodeLlama are sampled in a zero-shot setting as respective baselines. Higher is better. The darker the color, the better the score; heatmap is presented per each LLM and metric.}
    \label{tab:rq3-gen}
\end{table}

For evaluating code generation on R, we omit the CodeBLEU metric, as we could not access a readily available R tree-sitter parser to calculate the CodeBLEU score, and a custom implementation is out of the scope of this study. The results are summarized in Table~\ref{tab:rq3-gen}.

\textbf{T5 Family of Models.} Consistent with previous observations, fully fine-tuning the CodeT5 model achieves the highest Pass@1 and BLEU-4 score of $2.48$ and $14.20$, respectively. Closely following the fully fine-tuned CodeT5 model, LoRA achieves a Pass@1 score of $1.86$ and a BLEU-4 score of $13.15.$ IA$^3$ and Compacter rank lower than LoRA in terms of performance, with BLEU-4 scores of $9.02$ and $8.70$, respectively. Both IA$^3$ and Compacter obtain the same Pass@1 score of $0.62$. The EM@1 scores of all fine-tuning methods are arranged according to a similar pattern to the Pass@1 scores.
Though LoRA achieves close scores to fully fine-tuned CodeT5's BLEU-4 and Pass@1 scores, the low Pass@1 for all fine-tuning methods shows that the generated code is mostly non-executable for R.

Similar results are seen for the T5 model, where full fine-tuning has the highest BLEU-4, followed by LoRA. Compacter and IA$^3$ have lower scores. 
Similar to code generation on Python, T5 fails to pass any HumanEval problems using any fine-tuning technique, with all Pass@1 scores equal to zero. This suggests that T5 is not capable of learning any meaningful code generation abilities for R when tasked with generating out-of-distribution unseen code snippets.

Overall, for R code generation, all fine-tuning techniques obtain very low Pass@1 scores, indicating that T5 family models are not capable of generating executable code for R. In terms of generation textual similarity, LoRA is the best-performing fine-tuning technique, with IA$^3$ and Compacter following closely. However, the performance gap between LoRA and fully fine-tuned models is significantly larger for T5 compared to CodeT5, indicating that T5 benefits more from fully fine-tuning for R code generation.

\textbf{Llama Family of Models.}
For R code generation using CodeLlama, LoRA obtains the highest BLEU-4 of $26.30$, followed by Compacter ($24.81$) and IA$^3$ ($24.42$). The zero-shot setup, which uses no fine-tuning, lags behind significantly, yielding a BLEU-4 score of $8.22$. These results indicate that while fine-tuning boosts the model’s ability to generate textually similar code, there is still room for improvement in overall performance compared to Python with BLEU-4 scores of above $40$ for the PEFT techniques.

When evaluating CodeLlama on the HumanEval benchmark, the Pass@1 scores mirror the BLEU-4 results, with LoRA achieving the highest score at $20.50$. Compacter follows with a score of $18.01$, and IA$^3$ scores slightly lower at $17.39$. The zero-shot setup records a Pass@1 score of $11.18$, demonstrating that even without fine-tuning, CodeLlama can generate some executable R code. However, fine-tuning considerably improves its ability to solve R code tasks, particularly with LoRA.
Compared to Python using CodeLlama for code generation, the obtained Pass@1 scores are lower for R. 

Shifting the focus to the Llama2 model, performance is lower across both BLEU-4 and Pass@1 metrics compared to CodeLlama. LoRA remains the top-performing technique with a BLEU-4 score of $21.44$, followed by Compacter at $19.94$ and IA$^3$ at $19.60$. The zero-shot BLEU-4 score is lower than CodeLlama's, standing at $6.12$, which suggests that Llama2 is less capable of producing R code that aligns textually with the reference solutions when compared to its counterpart, CodeLlama.

For the HumanEval benchmark, Llama2’s performance is lower than CodeLlama's. LoRA achieves the highest Pass@1 score of $10.56$, followed by Compacter at $8.70$, and IA$^3$ at $7.45$. The zero-shot setting performs poorly, recording a Pass@1 score of $1.24$. These results highlight that Llama2 struggles more than CodeLlama to generate executable R code, even when fine-tuning techniques are applied. Similar results were observed for Python as discussed in Table \ref{tab:rq1-2-gen}.

Comparing the fine-tuning techniques on both Llama models, LoRA consistently delivers the highest BLEU-4 and Pass@1 scores, indicating that it is the most effective method for improving both text generation quality and executable code. Compacter and IA$^3$ follow closely in terms of performance, with Compacter slightly outperforming IA$^3$ in most cases. However, even with these fine-tuning techniques, the overall capability of both CodeLlama and Llama2 to generate executable R code remains limited.

Overall, for R code generation, the T5 family of models performs poorly in comparison to Llama models. CodeLlama achieves the best results among all models. While fine-tuning improves the models' ability to generate textually similar code and solve some test cases, none of the methods yield consistently strong results, except for the CodeLlama model. LoRA provides the best balance between textual similarity and code execution, but the gap between the models’ BLEU-4 and HumanEval Pass@1 scores suggests that generating functional R code remains a challenge for the Llama family.

\begin{tcolorbox}[customsummarybox]
LoRA exhibits the best R code generation performance followed closely by Compacter and IA$^3$ on the T5 and Llama family of models. Unlike Python code generation, the gap between the three is relatively small in the unseen code generation setting.

All three methods perform significantly on the code-specific models, with T5 failing to pass any of the HumanEval samples and Llama2 performing half as strongly as CodeLlama when assessed on HumanEval.
\end{tcolorbox}

\section{Discussions}\label{sec:discussions}
In this section, we discuss our findings. Due to time and computational resource restraints, and based on the obtained results, we only include code-LLMs, the code generation task and the Python and R languages (as \emph{seen} and \emph{unseen} languages) in the additional experiments.
Moreover, we omit IA$^3$ from all of the experiments conducted in the discussions to reduce time and computational resource requirements and avoid unreliable reports due to its inconsistent implementation, as explained in Section \ref{sec:study-design}.

\subsection{Computational Resource Efficiency}

\begin{table}
    \centering
    \setlength{\aboverulesep}{0pt}
    \setlength{\belowrulesep}{0pt}
    \resizebox{\textwidth}{!}{
        \begin{tabular}{l|c|c|c|c}
        \toprule
            \multirow{2}{*}{\textbf{PEFT}} &\multirow{2}{*}{\textbf{Total Param}} &\multirow{2}{*}{\textbf{Updated Param}} &\textbf{Peak Memory} &\textbf{Runtime} \\
            & & &\textbf{(GB)} &\textbf{(Minutes)} \\\midrule
            \multicolumn{5}{c}{\cellcolor[HTML]{d9d9d9}\emph{CodeT5}} \\
            \midrule
            Compacter &222,903,552 &0.051\% &2.97 &5.32 \\
            LoRA &222,903,552 &0.595\% &3.80 &5.53 \\
            \midrule
            \multicolumn{5}{c}{\cellcolor[HTML]{d9d9d9}\emph{T5}} \\
            \midrule
            Compacter &222,903,552 &0.051\% &2.97 &10.43 \\
            LoRA &222,903,552 &0.595\% &3.80 &8.45 \\
            \midrule
            \multicolumn{5}{c}{\cellcolor[HTML]{d9d9d9}\emph{CodeLlama}} \\
            \midrule
            Compacter &6,738,415,616 &0.012\% &13.99 &1.50 \\
            LoRA &6,738,415,616 &0.093\% &16.67 &1.56 \\
            \midrule
            \multicolumn{5}{c}{\cellcolor[HTML]{d9d9d9}\emph{Llama2}} \\
            \midrule
            Compacter &6,738,415,616 &0.012\% &13.99 &1.47 \\
            LoRA &6,738,415,616 &0.093\% &16.67 &1.57 \\
            \bottomrule
        \end{tabular}
    }
    \caption{The peak GPU memory and runtime of CodeT5, T5, CodeLlama and Llama2 fine-tuned with Compacter and LoRA, and their respective baselines during the training phase of the limited scenario. In this limited scenario, all models undergo $1000$ training steps with a batch size of $8$. The peak GPU memory usage and the runtime are reported in gigabytes and minutes, respectively. }
    \label{tab:resource}
\end{table}

While the performance of each PEFT technique is important, the resource usage is of significance too. We evaluate the peak memory usage and runtime of Compacter and LoRA in a controlled scenario where the training steps and the batch size are equal for all models to ensure fairness. In this setting, all models are trained on the Python code generation task for $1000$ steps with a train batch size of $8$.

Table \ref{tab:resource} illustrates the peak memory usage and runtime for training. For all models, Compacter utilizes a smaller GPU memory compared to LoRA.
Specifically, LoRA employs $27.9\%$ and $19.15\%$ more peak GPU memory on the T5 and Llama family of models, respectively. Moreover, Compacter completes the training phase in a slightly shorter period on all Llama family of models and CodeT5, when compared to LoRA. Only on T5, LoRA demonstrates a faster training phase.

\begin{figure}[ht]
    \centering
    \includegraphics[width=\textwidth]{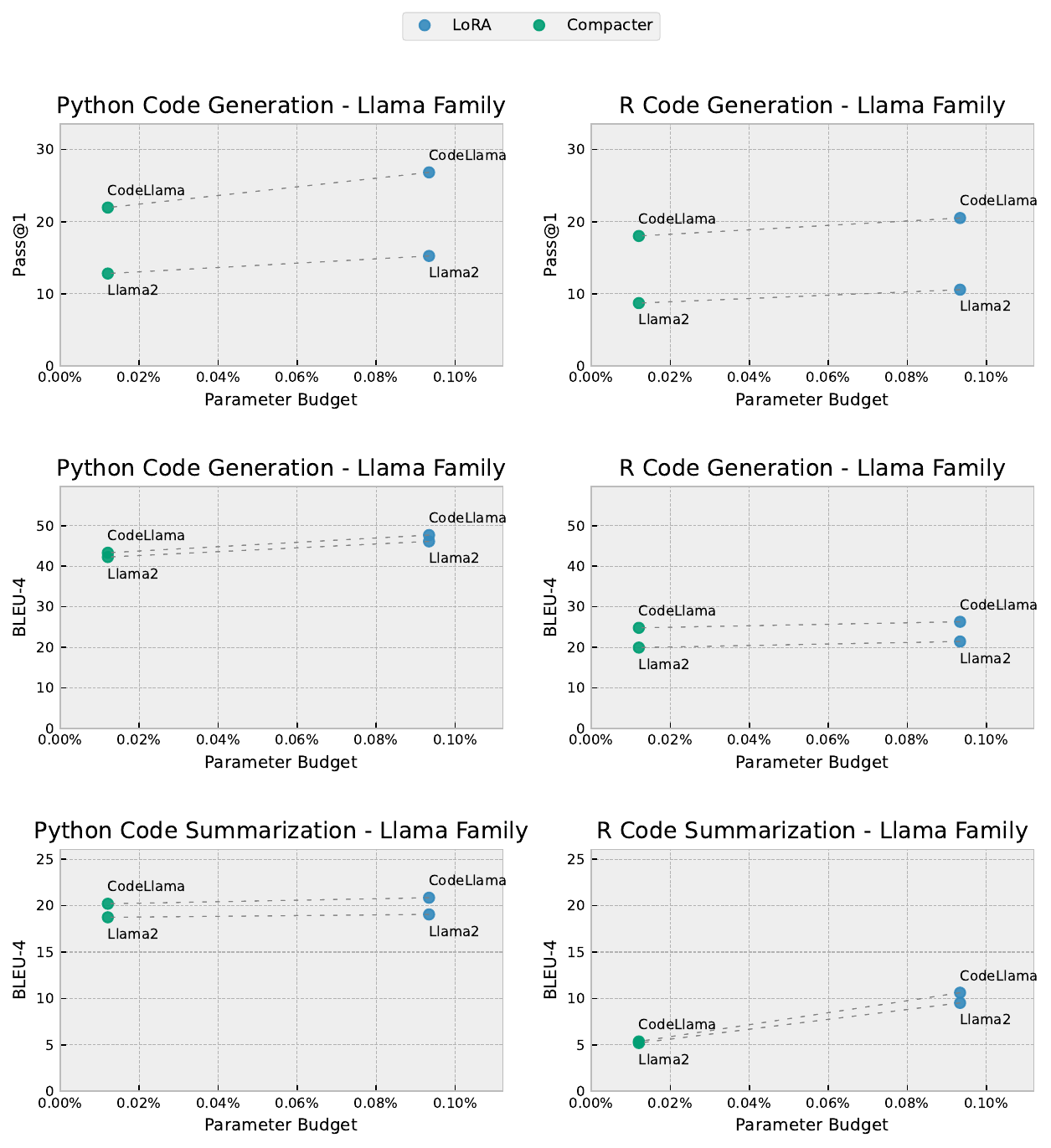}
    \caption{Performance versus parameter budget of Compacter and LoRA for Python and R code generation and summarization tasks using the Llama family of models. BLEU-4 represents performance. The parameter budget is the percentage of trained parameters compared to the number of total parameters.}
    \label{fig:res-Llama}
\end{figure}

\begin{figure}[ht]
    \centering
    \includegraphics[width=\textwidth]{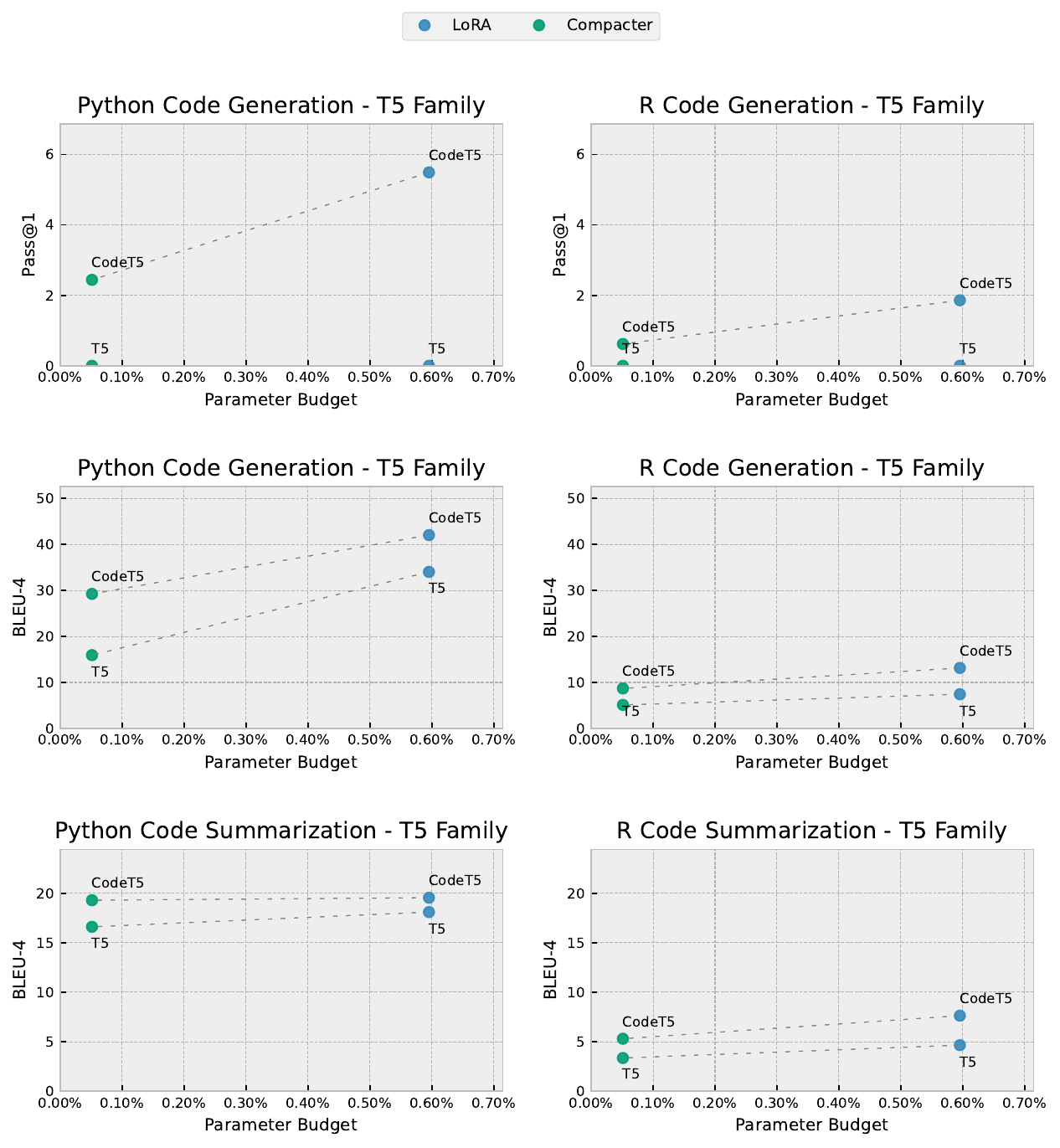}
    \caption{Performance versus parameter budget of Compacter and LoRA for Python and R code generation and summarization tasks using the T5 family of models. BLEU-4 represents performance. The parameter budget is the percentage of trained parameters compared to the number of total parameters.}
    \label{fig:res-T5}
\end{figure}

Besides resource requirements, performance is important in determining resource efficiency and guiding end-users in decision-making regarding performance-resource trade-offs. Given that Compacter's lower resource requirements can be attributed to its smaller trainable parameter budget the efficiency of both PEFT methods can be assessed by a comparative analysis between their trainable parameter budget and relative performance.

Figures \ref{fig:res-Llama} and \ref{fig:res-T5} depict Pass@1 (code generation) and BLEU-4 (code generation and code summarization) scores of LoRA and Compacter in Python and R for the Llama and the T5 family of models, respectively.
For R code generation and Python code summarization on Llama2 and CodeLlama, Compacter offers a relative performance close to LoRA's despite utilizing $7.75$ times less trainable parameters. However, this smaller trainable parameter budget may affect Compacter's capabilities for Python code generation and R code summarization, as LoRA pulls ahead on these tasks.

On the smaller T5 family of models, LoRA better illustrates its advantages. Besides Python code summarization where Compacter comes close to LoRA in terms of BLEU-4 score, LoRA outperforms Compacter dominantly for code generation in R and Python.

The larger resource requirements of LoRA become less effective on larger models. This is due to the fact that the trainable parameters introduced to the models constitute a larger portion of the total parameter count on smaller models (assuming similar hyperparameters for PEFT methods). For instance, while LoRA utilizes a $7.75$ times larger parameter budget on the Llama family of models compared to Compacter, this number increases to $11.66$ on the T5 family of models. The magnitude of this proportion directly affects the effectiveness of LoRA compared to Compacter.

Overall and across all models, Compacter undergoes training using less GPU memory and for most models completes the training phase faster when compared to LoRA. Even though its lower resource requirements come at the cost of worse performance (e.g., for code generation in R, Compacter achieved $18$ PAss@1 compared to $20.5$ Pass@1 by LoRA), Compacter can be a viable option depending on the base models and target tasks when available training resources become a limitation.

\begin{figure}[ht]
    \centering
    \includegraphics[width=\textwidth]{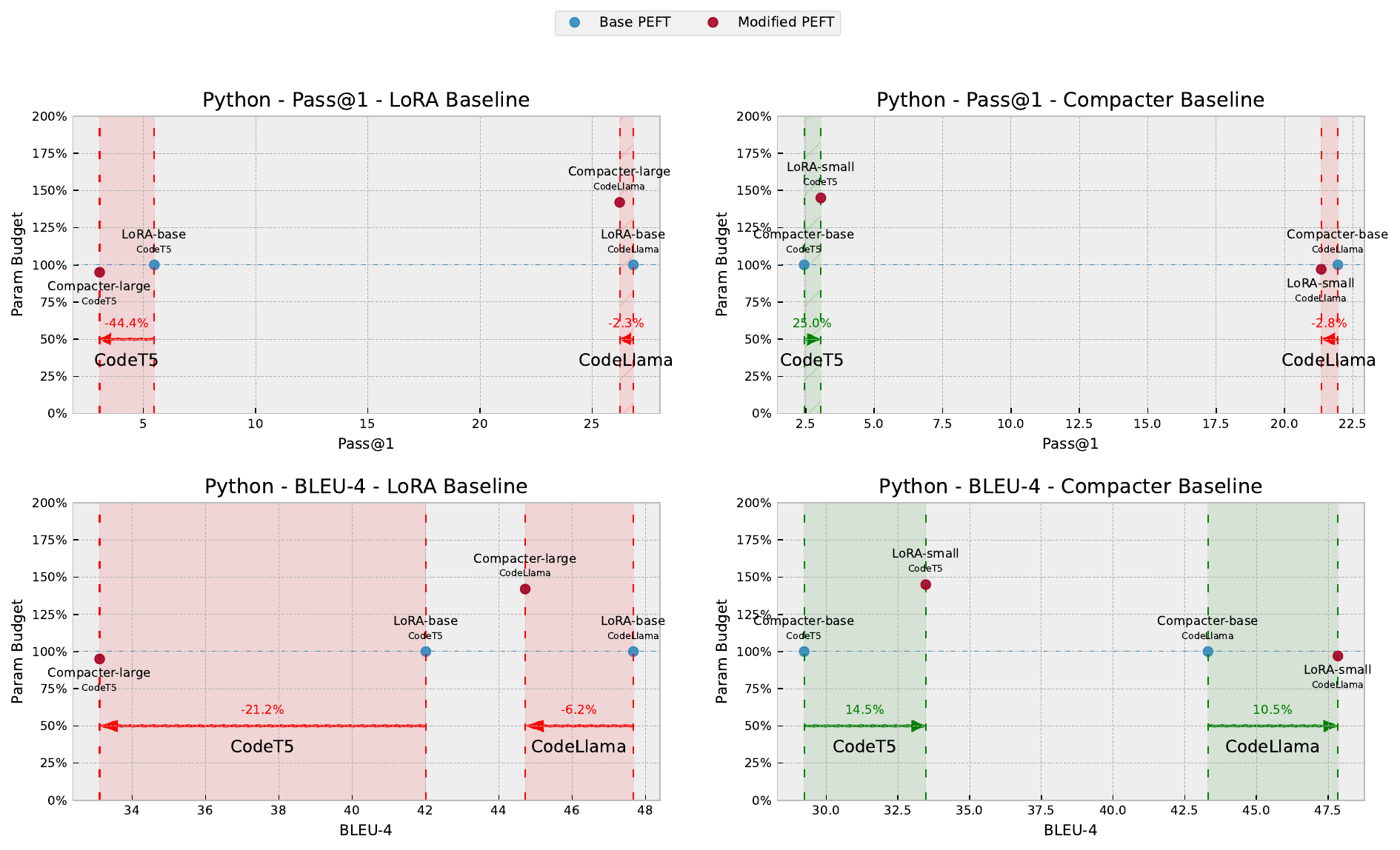}
    \caption{Performance difference of LoRA and Compacter on CodeT5 and CodeLlama for Python code generation, with various parameter budgets. Performance is represented by Pass@1 (top) and BLEU-4 (bottom), with a left-pointing arrow and a red area indicating a decrease and a right-pointing arrow and a green area indicating an increase in performance compared to the baseline PEFT. The parameter budget on the y-axis is the ratio of trainable parameters compared to the number of the baseline's trainable parameters.}
    \label{fig:same-param-python}
\end{figure}
\begin{figure}[ht]
    \centering
    \includegraphics[width=\textwidth]{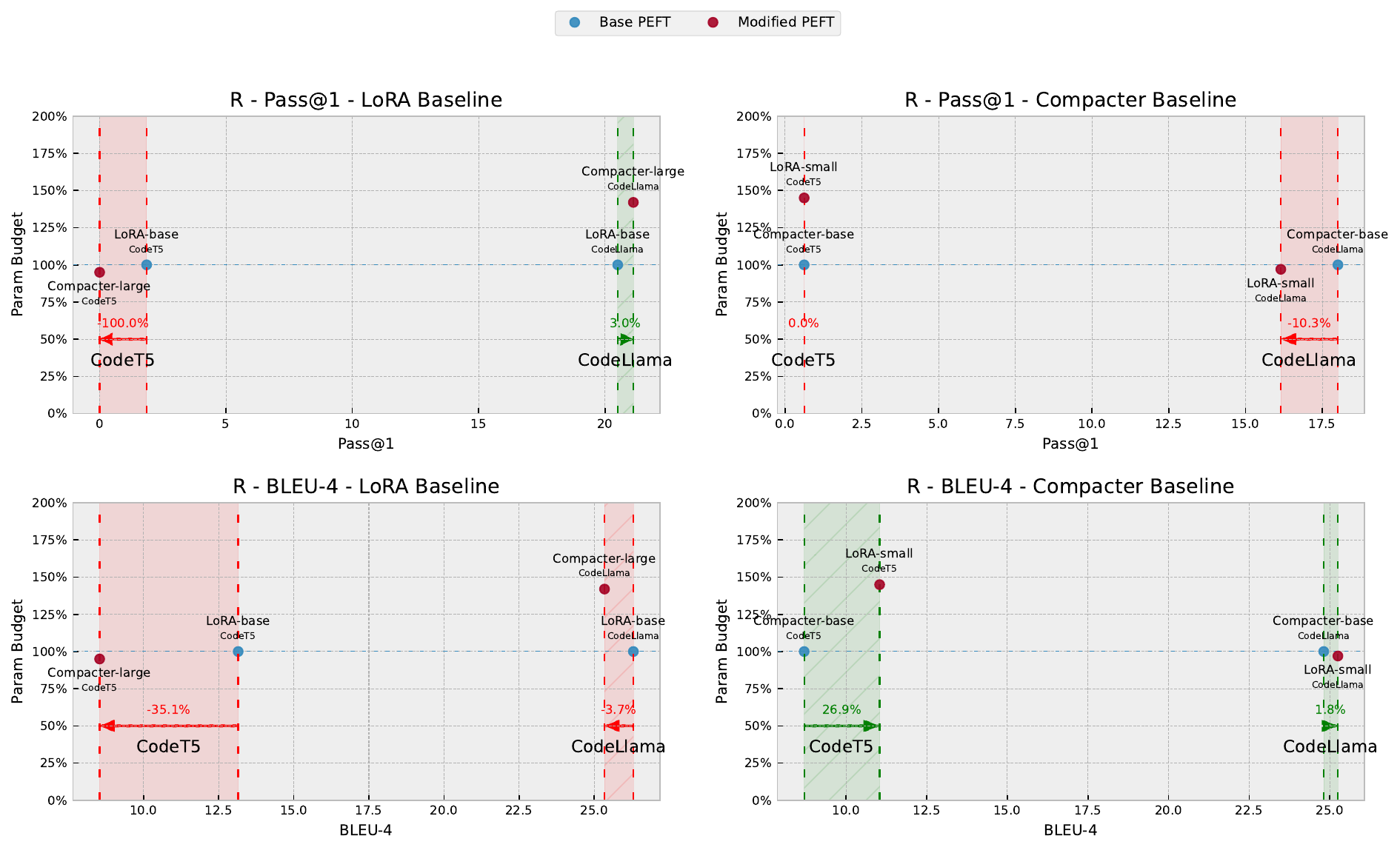}
    \caption{Performance difference of LoRA and Compacter on CodeT5 and CodeLlama for R code generation, with various parameter budgets. Performance is represented by Pass@1 (top) and BLEU-4 (bottom), with a left-pointing arrow and a red area indicating a decrease and a right-pointing arrow and a green area indicating an increase in performance compared to the baseline PEFT. The parameter budget on the y-axis is the ratio of trainable parameters compared to the number of the baseline's trainable parameters.}
    \label{fig:same-param-r}
\end{figure}

\subsection{Effect of Trainable Parameter Budget}

As shown previously, LoRA dominantly outperforms Compacter for Python and R. However, using the hyperparameters proposed by their authors, LoRA's increased performance comes at the cost of a larger trainable parameter budget and more computational resources. To compare their performance fairly, we modified the hyperparameters of both PEFT methods to reach the parameter budget of the other one. 
More specifically, in one scenario, we increased the parameter budget of Compacter to match the number of parameters used by LoRA, and in another, we decreased LoRA's parameters to become closer to the number of parameters of Compacter. In each case, we considered one of them as the baseline and the changed one as the modified one.

To achieve this, we used a higher $phm\_rank$ of $16$ on Compacter (referred to as Compacter-large) to increase its trainable parameter budget to $1,254,592$ and $8,921,152$ on CodeT5 and CodeLlama, respectively, and a lower $r$ of $1$ on LoRA (referred to as LoRA-small) to decrease its trainable parameter budget to $165,888$ and $786,432$ on CodeT5 and CodeLlama, respectively. 

We note that even though these parameter budget sizes are not exactly the same as the baseline parameter budget sizes (i.e., LoRA and Compacter default configurations in previous studies), given that we control discrete hyperparameters, they are within an acceptable range of the baselines relative to the much larger parameter budget of their base models. 
Due to computational restrictions, we only conduct this experiment for the code generation task. 
However, as we use the BLEU metric in addition to Pass@1, the textual similarity comparisons are considered here.

The general pattern we observe is that PEFT methods with similar parameter budgets obtain comparable performance in terms of functional accuracy (i.e., Pass@1) for both Python (top plots on Fig. \ref{fig:same-param-python}) and R (top plots on Fig. \ref{fig:same-param-r}). Compacter-large on CodeT5 receives the highest performance drop compared to the baseline LoRA for Python and R, followed by LoRA-small on CodeLlama for R. The Pass@1 score changes are minimal in other settings.

In terms of textual similarity (i.e., BLEU-4), however, we observe that LoRA is more robust in different parameter budgets (bottom plots on Fig. \ref{fig:same-param-python} for Python and Fig. \ref{fig:same-param-r} for R). When scaled down, LoRA-small consistently outperforms the baseline Compacter on both models and both languages, while the baseline LoRA outperforms the up-scaled Compacter-large in all settings.

Overall, we notice that for functional accuracy, i.e., Pass@1, the number of trainable parameters outweighs the effect of the PEFT's architecture. The architecture becomes more important for textual similarity, i.e., BLEU-4, where LoRA outperforms Compacter in similar parameter budgets.

\subsection{Robustness}

\begin{table}
    \centering
    \setlength{\aboverulesep}{0pt}
    \setlength{\belowrulesep}{0pt}

        \begin{tabular}{l|rrr|rrr}\toprule
        \multirow{2}{*}{\textbf{PEFT}} &\multicolumn{3}{c|}{\textbf{Python}} &\multicolumn{3}{c}{\textbf{R}} \\\cmidrule{2-7}
        &\textbf{Original} &\textbf{Modified} &\textbf{Change} &\textbf{Original} &\textbf{Modified} &\textbf{Change} \\
        \midrule
        \multicolumn{7}{c}{\cellcolor[HTML]{d9d9d9}\emph{CodeT5}} \\
        \midrule
        Compacter &11.99 &11.21 &-6.51\% \textcolor[HTML]{9d0000}{↓} &5.32 &5.78 &8.67\% \textcolor[HTML]{007212}{↑} \\
        LoRA &41.36 &37.84 &-8.52\% \textcolor[HTML]{9d0000}{↓} &13.13 &11.95 &-8.94\% \textcolor[HTML]{9d0000}{↓} \\
        \midrule
        \multicolumn{7}{c}{\cellcolor[HTML]{d9d9d9}\emph{T5}} \\
        \midrule
        Compacter &3.02 &2.71 &-10.47\% \textcolor[HTML]{9d0000}{↓} &4.22 &4.08 &-3.30\% \textcolor[HTML]{9d0000}{↓} \\
        LoRA &33.45 &31.74 &-5.11\% \textcolor[HTML]{9d0000}{↓} &8.36 &7.52 &-9.97\% \textcolor[HTML]{9d0000}{↓} \\
        \midrule
        \multicolumn{7}{c}{\cellcolor[HTML]{d9d9d9}\emph{CodeLlama}} \\
        \midrule
        Compacter &41.05 &41.54 &1.21\% \textcolor[HTML]{007212}{↑} &20.62 &20.37 &-1.19\% \textcolor[HTML]{9d0000}{↓} \\
        LoRA &48.68 &47.48 &-2.47\% \textcolor[HTML]{9d0000}{↓} &24.79 &22.65 &-8.62\% \textcolor[HTML]{9d0000}{↓} \\
        \midrule
        \multicolumn{7}{c}{\cellcolor[HTML]{d9d9d9}\emph{Llama2}} \\
        \midrule
        Compacter &40.13 &37.98 &-5.36\% \textcolor[HTML]{9d0000}{↓} &17.21 &15.55 &-9.67\% \textcolor[HTML]{9d0000}{↓} \\
        LoRA &46.00 &45.30 &-1.53\% \textcolor[HTML]{9d0000}{↓} &20.93 &20.01 &-4.41\% \textcolor[HTML]{9d0000}{↓} \\
        \bottomrule
        \end{tabular}

    \caption{The BLEU-4 score of all models fine-tuned using Compacter and LoRA for Python and R code generation tasks. The original column reports the BLEU-4 score across the unmodified 100 samples. The modified column reports the BLEU-4 score across the 100 samples modified to represent user typos and errors. The change column reports the change in the BLEU-4 score after modification.}
    \label{tab:robust}
\end{table}

For code generation, we also assess the robustness of PEFT methods in code generation tasks. Recent work by \citet{zhu2023promptbench} highlights that LLMs are prone to errors when prompted with inputs containing plausible user errors such as typos or synonyms. This emphasizes the need to understand how PEFT strategies can affect such challenges in code generation tasks. 
To achieve this, we randomly select 100 samples from Python and R, representing both seen and unseen languages in our experiments. The description docstrings within these samples are manipulated to mimic textual errors, i.e., user typos and improper synonyms, using GPT-4o\footnote{\url{https://platform.openai.com/docs/models\#gpt-4o}}. The prompt used to generate the modified samples and example original and modified prompts and their generations by each PEFT are present in Appendix \ref{apx:robust-prompt} and \ref{apx:robust-sample}, respectively.
We repeated the experiment three times and report the average scores in Table \ref{tab:robust}. 
By comparing the results obtained from these manipulated samples with their original format, we evaluate the robustness of Compacter and LoRA when applied to the T5 and Llama family of models. In this experiment, we compare performance using the BLEU-4 score.

As illustrated in Table \ref{tab:robust}, LoRA suffers from textual errors on all models for both languages. The performance loss for R is larger compared to Python, suggesting that the unseenness of R can impact LoRA's robustness. Notably, with LoRA, T5 and CodeT5 have the most performance decrease resulting from textual errors in R samples and Python samples, respectively. 
The Llama models have less performance decrease compared to the T5 models when using LoRA.

Compared to LoRA, Compacter tolerates textual errors better for both R and Python, in some cases even slightly surpassing the BLEU-4 score obtained by the original samples. 
In Python, textual errors do not impact the performance of any model fine-tuned with Compacter significantly. However, introducing textual errors to R samples results in a pattern where Compacter on code-LLMs, i.e., CodeT5 and CodeLlama, demonstrate strong robustness against these errors. On general-LLMs however, these textual errors can result in a notable performance drop.
A similar pattern is observed for Compacter for Python, resulting in $-1.4\%$ performance drop or $+3.3\%$ performance increase when applied on CodeT5 and CodeLlama, respectively. 

In summary, Compacter shows robustness in code generation when facing textual errors for all models, compared to LoRA. 
This observation might be due to having fewer parameters in Compacter, thus it can be less overfit to the data compared to LoRA. However, finding the exact reason requires a separate experiment, which is out of the scope of this study.

\subsection{Qualitative Study on Code Summarization Quality}
To further analyze the results obtained from the code summarization tasks by LoRA and Compacter, we conducted a qualitative analysis of the quality of the generated summaries.
For this task, we evaluated samples following the approach of previous studies~\cite{apicontext2com, lamner}, where we randomly selected $100$ samples from the test dataset for Python and R (other programming languages are eliminated due to resource restrictions), which contain the code, the ground truth and the generated summaries by the code-LLMs. 
Each sample is then ranked by two software engineers who are proficient in both programming languages according to three criteria based on a Likert scale of one to five for each criterion, where five is the highest. The criteria are as follows:

\begin{itemize}
    \item Informativeness (I), scores the comments according to covering the key aspects of code.
    \item Relevance (R), evaluates the consistency of the details in the comments compared to code.
    \item Fluency (F), scores the fluency of the comments in English, considering if the sentences are well written and are grammatically correct.
\end{itemize}

\begin{table}
    \centering
    \setlength{\aboverulesep}{0pt}
    \setlength{\belowrulesep}{0pt}

    \begin{tabular}{l|rrr|r|rrr|r}\toprule
        \multirow{2}{*}{\textbf{PEFT}} &\multicolumn{4}{c|}{\textbf{Python}} &\multicolumn{4}{c}{\textbf{R}} \\\cmidrule{2-9}
        &\textbf{I} &\textbf{R} &\textbf{F} &\textbf{Average} &\textbf{I} &\textbf{R} &\textbf{F} &\textbf{Average} \\
        \midrule
        \multicolumn{9}{c}{\cellcolor[HTML]{d9d9d9}\emph{CodeT5}} \\
        \midrule
        LoRA &3.89 &4.17 &\textbf{4.65} &4.24 &2.89 &3.2 &\textbf{4.15} &3.41 \\
        Compacter &\textbf{3.93} &\textbf{4.19} &4.62 &\textbf{4.25} &\textbf{3.12} &\textbf{3.3} &4.1 &\textbf{3.51} \\
        \midrule
        \multicolumn{9}{c}{\cellcolor[HTML]{d9d9d9}\emph{CodeLlama}} \\
        \midrule
        LoRA &\textbf{4.18} &4.36 &4.63 &4.39 &\textbf{3.55} &\textbf{3.54} &\textbf{4.06} &\textbf{3.72} \\
        Compacter &4.16 &\textbf{4.43} &\textbf{4.69} &\textbf{4.43} &3.39 &3.44 &4.04 &3.62 \\
        \midrule
        \bottomrule
        Ground Truth &4.68 &4.43 &4.35 &4.49 &4.14 &3.69 &4.01 &3.95 \\
        \bottomrule
    \end{tabular}
    \caption{The average scores given by the evaluators for Informativeness (I), Relevance (R) and Fluency (F) for LoRA and Compacter on CodeT5 and CodeLlama for Python and R code summarization. The last row represents the same scores for the ground truth samples of each language. All scores are out of five, with \textbf{bold} indicating the highest value in each row per language.}
    \label{tab:man_sum}
\end{table}

Table~\ref{tab:man_sum} represents the obtained scores across the three metrics for LoRA and Compacter on CodeLlama and CodeT5. The average score of both PEFT methods and both models are close to the ground truth, being slightly higher for CodeLlama. 
Among the three metrics, Informativeness is lower than other scores (R and F) for both Compacter and LoRA compared to the ground truth. Among the two models, CodeLlama has higher scores for all three metrics. 
Comparing Compacter and LoRA, we observe mixed results. On CodeT5, Compacter has higher scores in I and R, for both Python and R. 
However, on CodeLlama, LoRA has better scores in all three aspects for R but slightly lags behind Compacter in R and F for Python. 
Overall, both PEFT approaches have scores close to each other, especially for CodeLlama. 

For both PEFT methods on both base models, we observe a higher Fluency score compared to their respective ground truths, which may be due to the fact that these models have been pre-trained on large corpora of highly fluent data (in terms of grammar, punctuation and structure).
Beyond the small differences in scores, both PEFT methods are capable of reaching the performance of the relatively large CodeLlama on the much smaller CodeT5 model, suggesting adequate knowledge transfer on small code-LLMs for both Python and R.
For the R dataset, it is notable that the ground truth evaluations are around 4, out of 5, for all three aspects of informativeness, relevance, and fluency. This indicates that the quality of the datasets needs to be improved for R. 

\subsection{Qualitative Study on Styling Rules}

\begin{table}
    \centering
    \setlength{\aboverulesep}{0pt}
    \setlength{\belowrulesep}{0pt}
    \resizebox{\textwidth}{!}{
        \begin{tabular}{l|rrr|r|rrr|r}
        \toprule
        \multirow{2}{*}{\textbf{Category}} &\multicolumn{4}{c|}{\textbf{LoRA}} &\multicolumn{4}{c}{\textbf{Compacter}} \\\cmidrule{2-9}
        &Structure &Naming &Readability &\textbf{Total} &Structure &Naming &Readability &\textbf{Total} \\
        \midrule
        \multicolumn{9}{c}{\cellcolor[HTML]{d9d9d9}\emph{Python}} \\
        \midrule
        Passed &4.98 &4.95 &4.68 &14.61 &4.93 &4.98 &4.83 &\textbf{14.74} \\
        Failed &5.00 &4.95 &4.35 &\textbf{14.30} &4.68 &4.73 &4.85 &14.26 \\
        \midrule
        Average &4.99 &4.95 &4.51 &14.46 &4.80 &4.86 &4.84 &\textbf{14.50} \\
        \midrule
        \multicolumn{9}{c}{\cellcolor[HTML]{d9d9d9}\emph{R}} \\
        \midrule
        Passed &4.95 &4.95 &4.50 &14.40 &4.98 &4.93 &4.83 &\textbf{14.74} \\
        Failed &4.93 &4.98 &4.60 &\textbf{14.51} &4.68 &4.98 &4.68 &14.34 \\
        \midrule
        Average &4.94 &4.96 &4.55 &14.45 &4.83 &4.96 &4.75 &\textbf{14.54} \\
        \midrule
        \multicolumn{9}{c}{\cellcolor[HTML]{d9d9d9}\emph{Average}} \\
        \midrule
        Passed \& &\multirow{2}{*}{4.96} &\multirow{2}{*}{4.96} &\multirow{2}{*}{4.53} &\multirow{2}{*}{14.45} &\multirow{2}{*}{4.82} &\multirow{2}{*}{4.91} &\multirow{2}{*}{4.80} &\multirow{2}{*}{\textbf{14.52}} \\
        Failed & & & & & & & & \\
        \bottomrule
        \end{tabular}
    }
    \caption{The average structure, naming and readability (out of five) and the total scores (out of $15$) of LoRA and Compacter on CodeLlama for \emph{Passed} (i.e., samples that pass unit tests) and \emph{Failed} (i.e., samples that fail unit tests but are syntax error-free) for Python and R styling quality survey, reported by the participants. \textbf{Bold} in the total column indicates the highest score per row. The last row represents the average scores across \emph{Passed} and \emph{Failed} Python and R samples (i.e., average scores across both languages and categories).}
    \label{tab:style}
\end{table}

While the HumanEval benchmark has unit tests assessing the correctness of the generated code, it is incapable of assessing the styling rules of generated code. Following previous work \cite{cassano2023knowledge}, we will create one styling rubric for R and one for Python (other programming languages are omitted due to resource limitations) to access the capabilities of LoRA and Compacter in adhering to the best practices established by the software engineering communities. 
These rubrics contain three categories of rules based on Google's styling guidelines for R, and PEP 8 styling guidelines for Python. 

\begin{itemize}
    \item code structure and cleanness pertaining to the proper use of indentation, new lines, commenting and function definition,
    \item naming convention pertaining to appropriate naming of code constructs such as variables, functions, and
    \item code readability and best practices pertaining to the adoption of defined best practices such as usage of white spaces around operators.
\end{itemize}

Every category of rules will have an impact score on a scale of one to five according to the aforementioned guidelines. During the evaluation process, each code sample will receive a negative penalty for violating any of the rules in its corresponding rubric, based on the rule's impact score. Finally, the styling rating of each sample will be computed based on the accumulated penalties assigned to it. Rubrics we employed in this study are available in Appendix~\ref{apx:style-rubrics}.

This evaluation is conducted manually by two software engineers proficient in R and Python on $160$ randomly selected samples. The samples are chosen from HumanEval problems (i.e., unseen samples) to eliminate models' styling bias towards their training dataset. 
Upon initial processing, we observed that only CodeLlama is capable of generating code samples adequate for styling analysis, among which many do not pass as valid Python or R code snippets. To this end, we only evaluate samples generated by CodeLlama that have either passed their tests successfully or have failed them only due to logical errors (i.e., samples are syntactically correct). We separately evaluate the passed samples, named ``\emph{Passed}", and the failed samples, named ``\emph{Failed}", to maintain fairness in evaluation, as the performance of the model in terms of functional accuracy may affect the styling quality of the generated code snippets.
The inter-rater agreement among the two evaluators using Cohen-Kappa score \cite{cohenkappa} is $0.71$, which shows a high agreement among the two raters. 

\begin{figure}[ht]
    \centering
    \includegraphics[width=\textwidth]{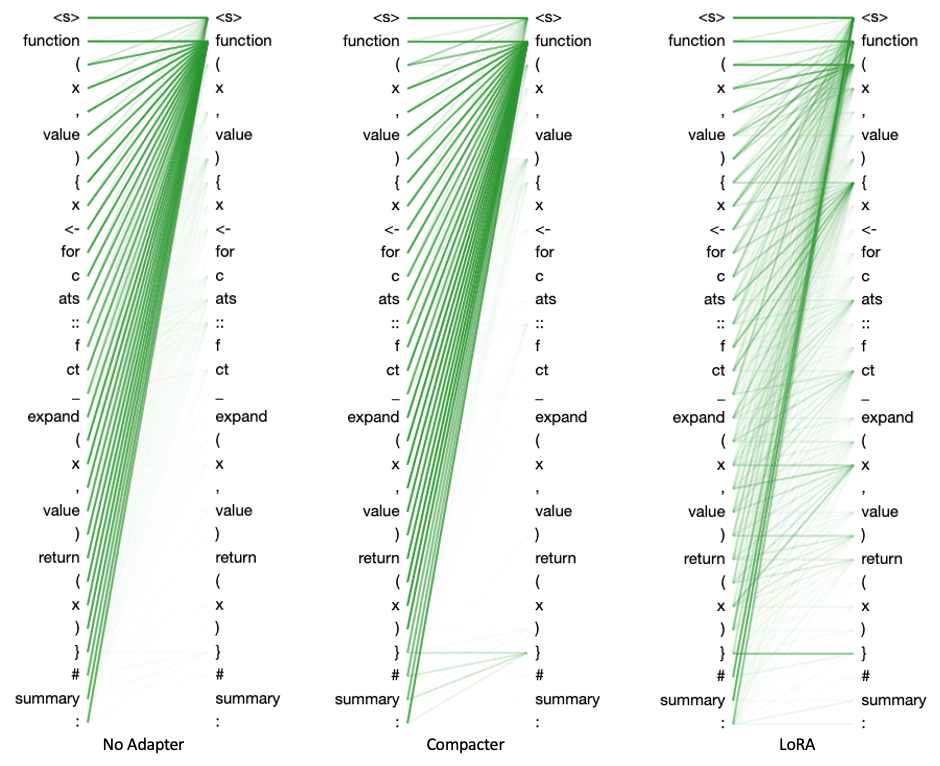}
    \caption{Attention map of CodeLlama to the input sequence on an example R function, with zero-shot setting (``No Adapter" in the plot), compared to the fine-tuned version of CodeLlama with Compacter and LoRA.}
    \label{fig:attention}
\end{figure}

As shown in Table \ref{tab:style}, while both PEFT methods follow the targeted styling rules very closely, overall Compacter tends to produce higher quality code in terms of styling rules for both the seen language (i.e., Python) and the unseen language (i.e., R)--see last row of Table \ref{tab:style}. Notably, however, LoRA exhibits better adherence to the styling rules when producing code that fails their unit tests (i.e., \emph{Failed} samples).
The overall high styling quality of generations produced by both PEFT methods may be due to the fact that the code samples in their training data already follow these rules, enabling both PEFT methods to generate code with the learned style. 

\subsection{Attention Changes}

Following previous works ~\cite{saberi2023utilization}, we highlight the attention change on an R code generation sample in the last transformer block of CodeLlama in the zero-shot setting and after fine-tuning with LoRA and Compacter in Fig. \ref{fig:attention}. We utilize Bertviz\footnote{\url{https://github.com/jessevig/bertviz}} to visualize the attention changes.

The attention map of the zero-shot CodeLlama (left plot of Fig. \ref{fig:attention}) suggests that the model 
mainly attends to the \emph{<S>} and \emph{function} tokens, without any significant attention to other tokens that construct the R code snippet. 
In contrast, both Compacter (middle plot of Fig. \ref{fig:attention}) and LoRA (right plot of Fig. \ref{fig:attention}) better attend to other tokens. 
Furthermore, we observe a large difference between the attention map of LoRA and Compacter, with the former attending to a more diverse set of important code tokens.

This difference in attention is aligned with the observed performance of these PEFT methods. However, the influence of attention map diversity and the performance exhibit a weak association; even though Compacter's attention map is closer to the zero-shot model, its performance is more comparable with LoRA's.

\section{Related Work}\label{sec:related} 
In this section, we review the works that study PEFT methods in Natural Language Processing and Software Engineering research. 

\subsection{PEFT Studies in Natural Language Processing}

\citet{low-res-related} conducted a comparative analysis of adapter-based fine-tuning and full fine-tuning across a diverse range of NLP tasks and datasets. Their findings highlighted several key observations. They found out that adapter-based tuning showcased superior performance in zero-shot cross-lingual tasks, and it exhibited higher stability and better generalization capabilities. Additionally, in scenarios requiring monolingual adaptation, adapter-based tuning demonstrated better outcomes in low-resource settings, particularly in more domain-specific tasks.
\citet{related-ding2023parameter} carried out extensive experiments involving numerous language models and various PEFT methods, including Adapters~\cite{adapter-houlsby2019parameter}, LoRA~\cite{lora-hu2022lora}, and Prefix-Tuning~\cite{prefix-li-liang-2021-prefix}, as well as their combination on a single model. Their study encompassed over 100 NLP tasks, examining these methods' performance, knowledge transferability, and scalability. 
\citet{related-lialin2023scaling} provided a detailed analysis of diverse PEFT methods introduced in recent years, categorizing them into three major classes: Addition-based, Selective-based, and Reparametrization-based. They conducted comparative assessments of these methods concerning storage efficiency, memory efficiency, computational efficiency, accuracy, and inference overhead.

\subsection{PEFT Studies in Software Engineering }

\citet{goel2022cross} was among the first to study adapters, a specific technique of PEFT, in software engineering. They used bottleneck adapters and assessed their capabilities for knowledge transfer from natural language models to code tasks. They extended the work in ~\cite{saberi2023utilization} to other software engineering tasks and code-LLM and showed that by using adapters, the model performance increases or is on par with fully fine-tuning results for the studied languages and tasks. 
\citet{ase2023} conducted an empirical study of PEFT methods for code summarization among other tasks using several encoder-only and encoder-decoder code-LLMs, including CodeBERT \cite{codebert-feng-etal-2020-codebert} and CodeT5 \cite{codet5-wang-etal-2021-codet5}. They investigated the effect of the number of trainable parameters on LoRA, adapter and two other PEFT methods, and their performance on low-resource, cross-lingual, and cross-projects.

\citet{weyssow2023exploring} investigated LoRA, IA$^3$ and Prompt Tuning \cite{prompt-lester-etal-2021-power} when applied to code-LLMs for code generation. Their primary emphasis lies in evaluating the benefits of PEFT methods when employed in large language models versus small language models. Additionally, they conducted a comparison between the performance of LLMs when combined with PEFT and In-Context Learning and provided more insights into the results.
\citet{tse2023} empirically evaluated various prompt tuning techniques for code intelligent tasks. They analyzed CodeBERT \cite{codebert-feng-etal-2020-codebert} and CodeT5 \cite{codet5-wang-etal-2021-codet5}, fully fine-tuned and with prompt tuning, for defect prediction, code summarization, and code translation in terms of accuracy and data efficiency.

\citet{rel1-Compre} evaluated fine-tuning of general-LLMs and code-LLMs including T5 and CodeT5 fine-tuned with several PEFT techniques such as LoRA, Houlsby \cite{adapter-houlsby2019parameter} and Pfeiffer \cite{pfeiffer-etal-2020-mad} adapters for clone and defect detection and code search and translation. They additionally analyzed the GPU usage and training time of these PEFT methods on the evaluated tasks.
In their empirical study, \citet{rel3-storhaug2024parameterefficientfinetuninglargelanguage} focused on unit test generation using CodeGen \cite{codegen-nijkamp2023codegen}, StarCoder \cite{li2023starcoder} and CodeLlama \cite{codellama-roziere2023code} family of models fine-tuned with LoRA, IA$^3$ and prompt tuning, and report that LoRA outperforms other studied PEFT methods, even surpassing full fine-tuning in some cases.

Though there are many studies using PEFT methods for code-related tasks, PEFT methods are under-researched in the low-resource domain and particularity for R. Nevertheless, \citet{cassano2023knowledge} proposed a novel technique to fine-tune LLMs on low-resource languages by generating semi-synthetic data from high-resource languages and leveraging LLM-generated unit tests. Using their approach, they evaluated the StarCoder and CodeLlama models on several low-resource languages including R before and after fine-tuning them on their datasets.
\citet{rel8-transfer} explored the transferability of several high-resource and low-resource languages for clone detection, error detection, solution domain classification and code repair using CodeT5. Building on their findings, they proposed a method to detect the best combination of source and target programming languages for knowledge transfer.
Though these works are related to low-resource languages, they do not study PEFT approaches.

The styling quality of LLM-generated code is relatively under-explored, particularly for PEFT methods and low-resource languages such as R.
\citet{rel11-shivashankar2024betterpythonprogrammingall} conduct an in-depth study on Python code quality generated by various language models, including language models fine-tuned with QLoRA \cite{dettmers2023qlora}.

\textbf{Differences between our work and the current literature.} 

The number of studies on PEFT approaches in software engineering is limited. With the advent of newer PEFT approaches, and large language models with billions of parameters, there is still a gap in understanding which of the NLP techniques can be applied to code, and to what extent. 
Though some of our proposed PEFT methods, target tasks, and LLMs have been used separately in previous works, our proposed study is novel and different in the following aspects.

First, to the best of our knowledge, no work delves into a direct comparison of the applicability of PEFT methods on code summarization and generation when applied to a diverse range of open-source code-LLMs and general-LLMs, including large-scale decoder-only models and relatively smaller encoder-decoder models. Open-source LLMs gained increased attention from the research community as they are accessible and provide opportunities for various research such as training the models, rather than just using the models. Therefore, we consider investigating open-source models an important aspect of our study.

Second, while Compacter is investigated by other works \cite{compacterrelated1-Hu2023ScaledPF, compacterrelated2-xu2024} in natural language, its application in the software engineering domain remains unknown. As far as we are aware, we are the first to employ and examine Compacter in software engineering.
Finally, employing PEFT for knowledge transfer to low-resource programming languages in software engineering is under-explored. R in particular, is a low-resource language that has not been studied in this domain before.

\section{Threats to Validity}\label{sec:threats} 
\textbf{Internal Threats.}
An internal threat can be associated with our chosen data types and quantization configuration for training and testing models. Since we conduct a comparative analysis and use identical settings for LoRA and Compacter, we avoid this threat in their case. For IA$^3$, we minimize the effect by only reporting its performance metrics and omitting it from further experiments in section \ref{sec:discussions}.

Hyperparameters and implementation could be the cause of another internal threat. We resort to the default hyperparameters of libraries and those employed by the authors of each PEFT method as the same best practice hyperparameters, in addition to exploring LoRA and Compacter's effectiveness with different hyperparameters in section \ref{sec:discussions}, and utilize greedy decoding to report deterministic and reproducible results. Given that implementation is out of this paper's scope, we resort to the implementations provided by well-established HuggingFace's peft \cite{peft-library} and AdapterHub \cite{adapterhub-poth2023adapters} libraries.

Another potential threat could be related to the evaluation of RQ3, where we assess the models' performance on the \emph{unseen} programming language, R. We have investigated the \emph{unseenness} by referring to the original papers of the selected models. CodeT5 has released its training data and R is not included in that dataset. Though the training dataset of CodeLlama is not public, the list of programming languages used in its training is available and R is not among them. So the threat related to R being seen during the pre-training of the models is alleviated.

The final internal threat revolves around analyzing the GPU memory utilization of the PEFT methods by measuring their peak memory usage, as libraries' internal implementations and functionalities such as caching mechanisms and periodical evaluation phases can introduce misleading momentary memory usage peaks. To alleviate this threat, we have taken measures to disable such functionalities. Additionally, for all models, we have confirmed the correctness of peak memory usage by comparing them against the average of periodical memory usage measurements during their training.

\textbf{External Threats.} 
This threat is related to the generalization of the results. We have conducted experiments for code summarization and code generation, two families of models, T5 and Llama, using IA$^3$, Compacter, and LoRA and unseen language R. Though we conducted several experiments, the results might not be applicable to other programming languages and PEFT methods, or other code-related tasks. 

\textbf{Construct Threats.} 
One thread can be our choice of models. We mitigate this threat by selecting models from a wide range of sizes, architecture and modalities (i.e., code and natural language LLMs) that are commonly employed in similar studies of language models in software engineering. Moreover, we omit specialized models (e.g., instruction-tuned models) and avoid prompt engineering practices to mitigate the confounding effects of prompting format on the effectiveness of PEFT methods.

Not opting to fully fine-tune the Llama family of models as baselines can be another threat. Given the substantial resources required, fully fine-tuning these models was not within our capabilities. We opt to use these models in zero-shot settings to minimize this threat, which is a common practice of leveraging large pre-trained LLMs in the literature.

A common construct threat in this domain is misrepresenting the overall performance of LLMs or PEFT methods by purely comparing metrics such as BLEU-4, CodeBLEU and Pass@k. We avoid this threat by clarifying how we applied each method and separately reporting these methods when available. Additionally, we perform a set of qualitative studies to shed light on other capability aspects of PEFT methods.

\textbf{Conclusion Threats.} 
Finally, the lack of sufficient experiment repetition can be considered a threat. Due to the significantly high number of experiments in this study, their resource cost and negative environmental side-effects, we avoided performing multiple runs of each fine-tuning. To mitigate this effect, however, we establish the significance of the results through inferential statistical tests, detailed in section \ref{sec:study-design}.
We also conducted several other experiments including manual evaluation and robustness to ensure the reliability of the results.

\section{Conclusion and Future Works}\label{sec:conclusion}
In this study, we presented an empirical comparison of PEFT methods within the domain of software engineering, employing a variety of foundational models, including code-specific and natural language LLMs. Our analysis encompasses the performance evaluation of each model augmented with distinct PEFT methods across six prevalent programming languages, accompanied by a comprehensive discussion of their effectiveness with different parameter budgets and their generation's characteristics.
Furthermore, we explore the PEFT methods' efficacy when applied to a programming language that was not included during the training phase, R in our case, to evaluate their performance in transferring knowledge to low-resource domains. 
This multifaceted analysis provides valuable insights into the adaptability and performance of PEFT methods in scenarios involving previously unseen programming languages, contributing to a deeper understanding of their practical applicability in software engineering contexts.
Future lines of research include an in-depth analysis of the characteristics of PEFT methods' architectures in software engineering to introduce more effective methods for code-related tasks as well as exploiting current techniques for automatic tools to support code generation and other code-related tasks in R.

\section*{Data Availability Statement}\label{sec:data-ack}
We include all scripts and tooling used to obtain the results in our GitHub repository.\footnote{\url{https://github.com/Amirresm/empirical_peft_suite}}

\section*{Conflict of Interest}
The authors declare that they have no conflict of interest.


\bibliographystyle{plainnat}
\bibliography{main}

\appendix
\section{Synthetic Dataset Prompt For Robustness Experiment} \label{apx:robust-prompt}

\lstset{
basicstyle=\tiny\ttfamily,
breaklines=true,
breakautoindent=false,
breakatwhitespace=false,
frame=single
}
    
\begin{lstlisting}[caption=Synthetic Dataset generation prompt,frame=t]{PLM}
Your job is to rewrite doc string to mimic a real-world user who is not proficient in english. Please rewrite the sentence to include misspellings, typos, and grammar mistakes. Then while keeping the errors, rephrase it so that it conveys the same idea but sounds different from the original version.
Only change the description of the function in the doc string. Do not change the example usecases in the doc string. Do not change the function signature, body or the name.
Only generate the modified doc string and function header. Do not add extra text. The output should have the same format as the input. 
Example input for Python:
def is_palindrome_string(test_str):
    """ Check if the given string is a palindrome or not.
    >>> is_palindrome_string("radar")
    True
    >>> is_palindrome_string("racecar")
    True
    >>> is_palindrome_string("raceca")
    False
    """

Example output for Python:
def is_palindrome_string(test_str):
    """ Is gave text palindorm?
    >>> is_palindrome_string("radar")
    True
    >>> is_palindrome_string("racecar")
    True
    >>> is_palindrome_string("raceca")
    False
    """


Example input for R:
# A quadratic cutoff that goes to zero smoothly at the cutoff boundary.
# Args:
#     r_cut (float): Cutoff value (in angstrom).
#     ri (float): Interatomic distance.
#     ci (float): Cartesian coordinate divided by the distance.
# Returns:
#     (float, float): Cutoff value and its derivative.
quadratic_cutoff <- function(r_cut, ri, ci) {

Example input for R:
# A curve drops down gradually to nothing right at the edge.
# Args:
#     r_cut (float): Cutoff value (in angstrom).
#     ri (float): Interatomic distance.
#     ci (float): Cartesian coordinate divided by the distance.
# Returns:
#     (float, float): Cutoff value and its derivative.
quadratic_cutoff <- function(r_cut, ri, ci) {


Input 
\end{lstlisting}

\section{Sample Generations of Robustness Experiment} \label{apx:robust-sample}
\lstset{
    language=Python,             
    basicstyle=\ttfamily\tiny,  
    keywordstyle=\color{blue},   
    stringstyle=\color{red},     
    commentstyle=\ttfamily\color[HTML]{188038},  
    numbers=left,                
    numberstyle=\tiny,           
    breaklines=true,             
    tabsize=4                    
}
\noindent\begin{minipage}[t]{.45\textwidth}
\begin{lstlisting}[caption=Python - LoRA - Original Sample,frame=t]{PLO}
def count_Dict(input_dict,key):
    """ Count the number of occurrences of the given key in a dictionary.
    >>> count_Dict({'a':1,'b':2,'c':3}, 'a')
    1
    >>> count_Dict({'a':1,'b':2,'c':3}, 'd')
    0
    >>> count_Dict({'a':1,'b':2,'c':3}, 'b')
    2
    """
    return (input_dict.get(key,0))
\end{lstlisting}
\end{minipage}\hfill
\begin{minipage}[t]{.45\textwidth}
\begin{lstlisting}[caption=Python - LoRA - Modified Sample,frame=t]{PLM}
def count_Dict(input_dict,key):
    """ How many time teh key apears in th dict?
    >>> count_Dict({'a':1,'b':2,'c':3}, 'a')
    1
    >>> count_Dict({'a':1,'b':2,'c':3}, 'd')
    0
    >>> count_Dict({'a':1,'b':2,'c':3}, 'b')
    2
    """
    count = 0
    for i in input_dict:
        if i == key:
            count += 1
    return count
\end{lstlisting}
\end{minipage}

\noindent\begin{minipage}[t]{.45\textwidth}
\begin{lstlisting}[caption=Python - Compacter - Original Sample,frame=t]{PCO}
def find_first_missing_positive(arr):
    """ Find the smallest positive number missing from the given array.
    >>> find_first_missing_positive([2, 1, 3])
    4
    >>> find_first_missing_positive([1, 2, 0])
    3
    >>> find_first_missing_positive([1, 2, 3, 4])
    5
    """
    arr.sort()
    for i in range(len(arr)):
        if arr[i] != i + 1:
            return i + 1
    return len(arr) + 1
\end{lstlisting}
\end{minipage}\hfill
\begin{minipage}[t]{.45\textwidth}
\begin{lstlisting}[caption=Python - Compacter - Modified Sample,frame=t]{PCM}
def find_first_missing_positive(arr):
    """ Finde smalest missin pozitive numbr in array give.
    >>> find_first_missing_positive([2, 1, 3])
    4
    >>> find_first_missing_positive([1, 2, 0])
    3
    >>> find_first_missing_positive([1, 2, 3, 4])
    5
    """
    for i in range(1, len(arr)+1):
        if i not in arr:
            return i
    return len(arr)+1
\end{lstlisting}
\end{minipage}

\lstset{
    language=R,             
    basicstyle=\ttfamily\tiny,  
    keywordstyle=\color{blue},   
    stringstyle=\color{red},     
    commentstyle=\ttfamily\color[HTML]{188038},  
    numbers=left,                
    numberstyle=\tiny,           
    breaklines=true,             
    tabsize=4                    
}
\noindent\begin{minipage}[t]{.45\textwidth}
\begin{lstlisting}[caption=R - LoRA - Original Sample,frame=t]{RLO}
#Assert that actual_value is within range (inclusive).
range <- function(actual_value, lower_limit, higher_limit) {
  if (actual_value < lower_limit || actual_value > higher_limit) {
    stop(
      sprintf(
        "Expected value to be between %s and %s, but was %s",
        lower_limit, higher_limit, actual_value
      )
    )
  }
  return(actual_value)
}
\end{lstlisting}
\end{minipage}\hfill
\begin{minipage}[t]{.45\textwidth}
\begin{lstlisting}[caption=R - LoRA - Modified Sample,frame=t]{RLM}
# Check actual_vale inside the l o h limits.
range <- function(actual_value, lower_limit, higher_limit) {
  if (actual_value >= lower_limit && actual_value <= higher_limit) {
    return(TRUE)
  } else {
    return(FALSE)
  }
}
range(10, 5, 15)
range(10, 15, 5)
range(10, 15, 15)
range(10, 5, 10)
range(10, 10, 5)
range(10, 5, 10)
range(10, 10, 10)
range(10, 10, 15)
range(10, 15, 10)
range(10, 15, 15)
range(10, 15, 10)
range(10, 15, 15)
range(10, 15, 15)
range(10, 15, 15)
range(10, 15, 15)
range(10, 15
\end{lstlisting}
\end{minipage}

\noindent\begin{minipage}[t]{.45\textwidth}
\begin{lstlisting}[caption=R - Compacter - Original Sample,frame=t]{RCO}
#  Extracts all digits from incomig string 
get_num <- function(x) {
  x <- gsub("[^0-9]", "", x)
  return(as.numeric(x))
}
\end{lstlisting}
\end{minipage}\hfill
\begin{minipage}[t]{.45\textwidth}
\begin{lstlisting}[caption=R - Compacter - Modified Sample,frame=t]{RCM}
# Pulls out all nums from givven text 
get_num <- function(x) {
  nums <- c()
  for (i in 1:length(x)) {
    if (is.numeric(x[i])) {
      nums <- c(nums, x[i])
    }
  }
  return(nums)
}
\end{lstlisting}
\end{minipage}
\section{Styling Rubrics} \label{apx:style-rubrics}
\lstset{
    language=Python,             
    basicstyle=\ttfamily\small\color[HTML]{188038},  
    keywordstyle=\color{blue},   
    stringstyle=\color{red},     
    commentstyle=\color{green},  
    numbers=left,                
    numberstyle=\tiny,           
    breaklines=true,             
    tabsize=4                    
}

\subsection{Python Styling Rubric}
\begin{enumerate}
    \item \textbf{Code Structure and Cleanness (5 points)}
    \begin{itemize}
        \item \textbf{Unused comments or Code:}
        \begin{itemize}
            \item Comments must provide information or elaborate. Generation must be clear of comments that provide no additional information. 
            \item Generated code must be limited to the target function. Generation must be clear of code snippets written outside of the target function.
            \item Deduct points for unnecessary comments and code snippets separately.
        \end{itemize}
        \item \textbf{Indentation and Line Length:}
        \begin{itemize}
            \item Ensure code uses \textbf{4 spaces} per indentation level (not tabs).
            \item Nested code blocks (e.g., within loops or if statements) must be properly aligned with consistent indentation.
            \item Deduct points for inconsistent indentation within the same function, such as mixing tabs and spaces or misaligned code blocks.
            \item Each line should be limited to a maximum of \textbf{79 characters}.
            \item Ensure that long expressions are broken into multiple lines, with proper continuation indentation (use 8 spaces for continuation or align with the opening delimiter of parentheses).
            \item Deduct points for unnecessarily long lines or improper line breaks.

        \end{itemize}
        \item \textbf{Blank lines and Bracket placement:}
        \begin{itemize}
            \item Ensure logical sections within functions are separated by blank lines (e.g., separating the setup of variables from logic processing).
            \item Blank lines should not be overused. One blank line suffices between separate sections in a function.
            \item Ensure proper placement of parentheses and brackets for readability (e.g., open brackets should align with the first non-whitespace character of the line for line continuations).

        \end{itemize}
    \end{itemize}

    \item \textbf{Naming Conventions (5 points)}
    \begin{itemize}
        \item \textbf{Variable Names:}
        \begin{itemize}
            \item Use snake\_case for variable names (e.g.,  \lstinline[columns=fixed]{data_frame}, not \lstinline[columns=fixed]{DataFrame} or \lstinline[columns=fixed]{dataFrame}).
            \item Variable names should be \textbf{descriptive and meaningful}, reflecting their purpose in the function. Avoid single-letter variables except in trivial cases (e.g., loop indices like  \lstinline[columns=fixed]{i},  \lstinline[columns=fixed]{j}).
            \item Deduct points for using non-descriptive or unclear names, or for inconsistent use of naming conventions.

        \end{itemize}
        \item \textbf{Function Names:}
        \begin{itemize}
            \item Ensure function and argument names are in snake\_case and clearly reflect what the function does.
            \item For functions with side effects, names should indicate the action being performed (e.g., \lstinline[columns=fixed]{calculate_sum()} or \lstinline[columns=fixed]{print_data()}).
            \item Deduct points for unclear or overly generic function names (e.g., \lstinline[columns=fixed]{do_stuff()}), or inconsistent use of naming conventions within the same codebase.
        \end{itemize}
    \end{itemize}
    
    \item \textbf{Code Readability and Best Practices (5 points)}
    \begin{itemize}
        \item \textbf{Use of spaces:}
        \begin{itemize}
            \item Ensure there is a consistent and appropriate use of spaces around operators (e.g., \lstinline[columns=fixed]{x + y}, not \lstinline[columns=fixed]{x+y}).
            \item Inside parentheses, brackets, or braces, avoid extraneous spaces (e.g., \lstinline[columns=fixed]|foo(x[1], {bar: 2})| instead of \lstinline[columns=fixed]|foo( x[ 1 ], { bar : 2 } )|).
            \item Deduct points for inconsistent or improper use of spaces that reduce readability.
        \end{itemize}
        \item \textbf{Use of idiomatic Python:}
        \begin{itemize}
            \item Prefer Pythonic constructs, such as list comprehensions, where appropriate.
            \item Use built-in functions like \lstinline[columns=fixed]{sum()}, \lstinline[columns=fixed]{any()}, or \lstinline[columns=fixed]{max()} instead of loops where applicable.
            \item Ensure loops are used efficiently, avoiding unnecessary computations.
            \item Deduct points for overly complex or non-idiomatic solutions that could be simplified with standard Python features.
        \end{itemize}
    \end{itemize}
\end{enumerate}

\subsection{R Styling Rubric}
\begin{enumerate}
    \item \textbf{Code Structure and Cleanness (5 points)}
    \begin{itemize}
        \item \textbf{Unused comments or Code:}
        \begin{itemize}
            \item Comments must provide information or elaborate. Generation must be clear of comments that provide no additional information. 
            \item Generated code must be limited to the target function. Generation must be clear of code snippets written outside of the target function.
            \item Deduct points for unnecessary comments and code snippets separately.
        \end{itemize}
        \item \textbf{Indentation and Line Length:}
        \begin{itemize}
            \item Code must be consistently indented with \textbf{2 spaces per level} (not tabs).
            \item Nested code blocks (e.g., inside if statements or loops) should maintain consistent indentation.
            \item Deduct points for inconsistent indentation, such as mixing tabs and spaces or misaligned code blocks.
            \item Limit each line to a maximum of \textbf{80 characters} to maintain readability.
            \item For long expressions, break the code into multiple lines and align with proper indentation.
            \item Ensure line breaks occur at logical points (e.g., after a comma, or before an operator).
            \item Deduct points for overly long lines or poor line-breaking practices.
        \end{itemize}
        \item \textbf{Blank lines and Bracket placement:}
        \begin{itemize}
            \item Use blank lines to separate logical sections within the function (e.g., separating setup, processing, and return).
            \item Avoid excessive or inconsistent use of blank lines (one blank line between distinct sections suffices).
            \item Place the opening brace \{ on the same line as the control statement (e.g., \lstinline[columns=fixed]{if}, \lstinline[columns=fixed]{for}), with the closing brace aligned with the control statement.
            \item Deduct points for inconsistent brace placement or misaligned closing braces.
        \end{itemize}
    \end{itemize}

    \item \textbf{Naming Conventions (5 points)}
    \begin{itemize}
        \item \textbf{Variable Names:}
        \begin{itemize}
            \item Variable names should follow the snake\_case convention (e.g., \lstinline[columns=fixed]{data_frame}, not \lstinline[columns=fixed]{dataFrame} or \lstinline[columns=fixed]{DataFrame}).
            \item Names must be \textbf{descriptive and meaningful}, reflecting their role in the function. Avoid single-character names except for loop counters (e.g., \lstinline[columns=fixed]{i}, \lstinline[columns=fixed]{j}).
            \item Deduct points for non-descriptive, vague, or inconsistent naming practices.
        \end{itemize}
        \item \textbf{Function Argument Names:}
        \begin{itemize}
            \item Function names and arguments should use snake\_case and be \textbf{meaningful}, clearly reflecting their role in the function (e.g., \lstinline[columns=fixed]{user_data}, \lstinline[columns=fixed]{file_path}).
            \item Avoid abbreviations unless they are widely understood in context.
            \item Deduct points for ambiguous or overly abbreviated argument names.
        \end{itemize}
    \end{itemize}
    
    \item \textbf{Code Readability and Best Practices (5 points)}
    \begin{itemize}
        \item \textbf{Spacing:}
        \begin{itemize}
            \item Use proper spacing around operators for readability (e.g., \lstinline[columns=fixed]{x <- y + 2}, not \lstinline[columns=fixed]{x<-y+2}).
            \item Inside function calls and indexing, avoid extraneous spaces (e.g., \lstinline[columns=fixed]{foo(x[1], y)} instead of \lstinline[columns=fixed]{foo( x[ 1 ], y )}).
            \item Deduct points for inconsistent or improper spacing that reduces readability.
        \end{itemize}
        \item \textbf{String Handling:}
        \begin{itemize}
            \item \textbf{Single quotes} (\lstinline[columns=fixed]{'}) should be used for text where possible, unless there’s a compelling reason to use double quotes (e.g., for embedding quotes inside a string).
            \item Deduct points for improper or inconsistent use of quotes for strings.
        \end{itemize}
        \item \textbf{Explicit Return:}
        \begin{itemize}
            \item Functions should explicitly return a value using the \lstinline[columns=fixed]{return()} function, even if the result is the last evaluated expression.
            \item Deduct points for relying on implicit returns, as it reduces clarity.
        \end{itemize}
    \end{itemize}
\end{enumerate}

\end{document}